\documentclass[letter,aps,pra,amsmath,amssymb,reprint,showkeys,showpacs]{revtex4-1} 

\usepackage{graphicx}
\usepackage{subfigure}
\usepackage{lipsum}
\usepackage{braket}
\usepackage{dcolumn}

\newcommand{\unit}[1]{\ensuremath{\, \mathrm{#1}}}

\usepackage{bm}
\newcommand{\vect}[1]{\boldsymbol{\mathbf{#1}}}
\newcommand{\identity}{1\!\!1}

\begin{document}

\title{Reduction of the dc electric field sensitivity of circular Rydberg states using non-resonant dressing fields}
\author{Y.~Ni}
\author{P.~Xu}
\author{J.~D.~D.~Martin}
\affiliation{Department of Physics and Astronomy and Institute for Quantum Computing, University of Waterloo, Waterloo, Ontario, Canada N2L 3G1}
\date{\today}

\begin{abstract}
Non-resonant dressing fields can make the transition frequency between two circular Rydberg states insensitive to second order variations in the dc electric field.  Perturbation theory can be used to establish the required dressing field amplitude and frequency.  The same perturbative approach may be used to understand removal of the first order dependence of the transition frequency on electric field about a bias dc electric field [Hyafil {\it et al.}~Phys.~Rev.~Lett.~{\bf 93}, 103001 (2004)].  The directional alignment of the dressing and dc fields is critical in determining the electric field sensitivity of the dressed transition frequencies.  This sensitivity is significantly larger for circular Rydberg states compared to low-angular momentum Rydberg states of Rb.
\end{abstract}


\maketitle

\section{Introduction}

Large electric transition dipole moments can exist between Rydberg states of comparable energies.  These dipole moments enable strong coupling to electromagnetic fields at rf/microwave frequencies \cite{raimond:2001} and enhance interactions between Rydberg atoms \cite{marcassa:2014}.  Large electric transition dipole moments also lead to enhanced dc polarizabilities of Rydberg states --- the dc Stark shifts of low angular-momentum (low-$\ell$) Rydberg states scale with principal quantum number $n$ like $n^7$ \cite{gallagher:1994}.

The strong response of Rydberg atom energy levels to dc electric fields can be useful, such as for the acceleration and deceleration of the atomic center of mass \cite{breeden:1981,*procter:2003,*vliegen:2004}, but in other situations this sensitivity is a nuisance \cite{sandoghdar:1996,*weidinger:1997,hogan:2011,*thiele:2014}.  For example, as a surface is approached --- where there are inhomogeneous dc and low-frequency fluctuating electric fields --- it may be desirable to maintain well-defined Rydberg-Rydberg transition frequencies for resonant coupling to surface devices \cite{sorensen:2004,*petrosyan:2009, *pritchard:2014}.  The large transition dipole moments of Rydberg atoms enable strong resonant coupling, but they also lead to troublesome dc Stark shifts, spoiling the resonance condition.

With the motivation of creating robust qubits, a variety of approaches have been taken in different physical systems to make resonance frequencies less sensitive to external perturbations. Significant progress has been made in superconducting qubits  \cite{vion:2002} and trapped ion qubits \cite{timoney:2011}, to name but two examples.  In the case of ground electronic state neutral atoms, non-resonant ``dressing'' fields have been shown to reduce the sensitivity of hyperfine transition frequencies to magnetic field fluctuations \cite{zanon:2012,*sarkany:2014}.

It is also desirable to reduce the sensitivity of Rydberg-Rydberg transition frequencies to low-frequency and dc electric fields while maintaining their high sensitivity to resonant fields.  Hyafil {\it et al.}~\cite{hyafil:2004} proposed the use of non-resonant dressing fields to eliminate the first order dependence of the transition energy between two circular ($|m|=\ell=n-1$) Rydberg states on dc field for fluctuations around a non-zero ``bias'' dc electric field (further details are given in Ref.~\cite{mozley:2005}).  Bason {\it et al.}~\cite{bason:2010} used non-resonant dressing fields to {\em increase} the sensitivity of Rydberg state energies to dc electric fields, and Sevin\c cli and Pohl \cite{sevincli:2014} have explored the influence of multiple dressing fields on Rydberg-Rydberg interactions.  Recently, Jones {\it et al.}~\cite{jones:2013} experimentally demonstrated a significant reduction in the electric field dependence of low-$\ell$ Rydberg-Rydberg transition frequencies through the application of a non-resonant dressing field.

\begin{figure}
\includegraphics{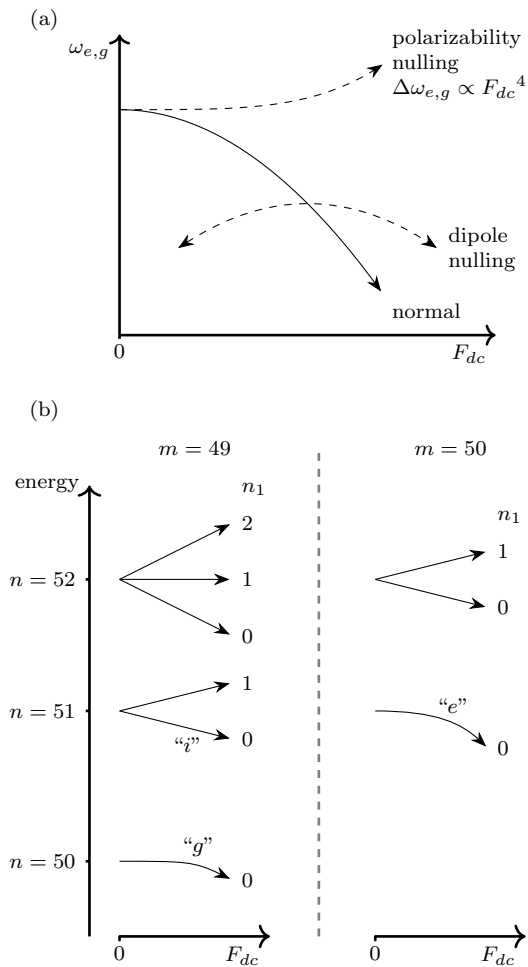}
\caption{\label{fg:schematic}
(a)  Polarizability and dipole-nulling contrasted with the normal dc electric field dependence of a transition frequency $\omega_{e,g}$. (b) schematic energy level structure near the two circular states $e$ and $g$ ($n_1$ is the parabolic quantum number \cite{bethe:1957}).   This diagram is purely qualitative; for example, under the conditions studied here, the Stark shifts are much smaller than the energy separation between states of consecutive $n$.
}
\end{figure}

In the work of Jones {\it et al.}~\cite{jones:2013}, the reduction in electric field sensitivity was discussed in terms of the elimination of the first-order dependence (dipole) of the dressed energy level difference on electric field about a specific non-zero dc field: so-called {\em dipole nulling} (see Fig.~\ref{fg:schematic}a).  However, the experimental results showed a stronger suppression of the variation of transition energy with dc electric field.  More specifically, the residual deviations of the transition energy were observed to be quartic with dc electric field rather than the quadratic behavior expected for dipole nulling \cite{hyafil:2004,mozley:2005}.  In effect, the experimental spectra were {\em polarizability nulled} (see Fig.~\ref{fg:schematic}a).  In contrast, the calculations of Hyafil and Mozley {\it et al.}~\cite{hyafil:2004,mozley:2005} for circular Rydberg states exhibited only dipole-nulling.

In the present study, we establish that polarizability nulling can be achieved for circular states and compare this with the low-$\ell$ Rb case \cite{jones:2013}.

Understanding dressed polarizabilities and dipole moments for circular Rydberg atoms is significant, because circular states are both experimentally realizable and have special properties such as long radiative lifetimes \cite{hulet:1983}.  All atomic species have quantitatively similar circular Rydberg states due to the low core penetration of high-$\ell$ states.  As shall be shown, the energy degeneracy of high-$\ell$ states gives nulling behavior qualitatively distinct from that of low-$\ell$ states,
particularly when the dressing and dc electric fields are not parallel in space.

\section{Techniques}
\label{se:techniques}

\subsection{Dipole matrix elements}

To understand the influence of dressing fields on the dc electric field response of Rydberg atoms, we consider the zero field Hamiltonian $H_0$ with electric dipole couplings:
\begin{equation}
H = H_0 
- \vect{\mu} \cdot \vect{F}_{\rm dc} 
- \vect{\mu} \cdot 
\left[(\vect{F}_{\rm ac}/2) \exp ( i \omega_d t ) + 
{\rm c.c.} \right],
\label{eq:basic_hamiltonian}
\end{equation}
where $\vect{\mu}$ is the electric dipole operator and $\vect{F}_{\rm dc}$ is the dc electric field.  The dressing field of angular frequency $\omega_d$ is described using a complex amplitude vector $\vect{F_{ac}}$ which is convenient for the consideration of elliptically polarized dressing fields \cite{jones:2013}.  

It is assumed that the ionic core of the Rydberg atom is in a single state independent of the state of the outer electron, and that we need only consider the wavefunctions of the outer (Rydberg) electron to calculate the electric dipole couplings.  To compactly describe the states of the Rydberg electron, we use the approach of Zimmerman {\it et al.}~\cite{zimmerman:1979}, who consider as a basis set for the dc Stark effect, a set of bound states of the Rydberg atom, with energies set by the spectroscopy of the zero field Rydberg series.  With electric fields that produce insignificant ionization rates, the continuum states can be neglected.

In the case of circular Rydberg states --- where the spin-orbit splitting is insignificant --- the basis states are labelled as $\ket{n \ell m}$ where $n$ is the principal quantum number of the Rydberg electron, $\ell$ is its angular momentum, and $m$ is the projection of this angular momentum on the $z$ axis.

Due to their limited core penetration, high-$\ell$ states (such as the circular states) are approximated using hydrogenic wavefunctions.  To compute the radial parts of the matrix elements between these states, we use Eq.'s (63.2) and (63.5) of Ref.~\cite{bethe:1957}.  (For these two equations to have consistent radial wavefunction phase definition, a minus sign should be placed in front of Eq.~(63.5).)

For the low-$\ell$ states of Rb, spin-orbit coupling cannot be neglected.  Instead we use basis states labeled as $\ket{n\ell jm_j}$ where $j$ and $m_j$ arise from the operator $\vect{j} = \vect{\ell} + \vect{s}$, where $\vect{s}$ refers to the unpaired electron spin.  Electric dipole couplings between the low-$\ell$ states of Rb are computed in an identical manner to Ref.~\cite{zimmerman:1979}: the radial Rydberg wavefunctions are integrated in from large $r$, assuming a purely Coulombic potential, with a zero-field energy
 $E_{n\ell j}$ (i.e.~$H_0 \ket{n \ell j m_j} = E_{n \ell j} \ket{n \ell j m_j}$) 
based on the known Rb Rydberg state energy levels \cite{li:2003}.  Beneath a certain low-$r$ the integration is terminated and the wavefunction assumed to be zero at lower $r$.  Wavefunctions obtained in this manner can be used to compute 
the radial contribution to 
$\bra{n' \ell' j' {m_j}'}\vect{\mu}\ket{n \ell j m_j}$.  
Evaluation of the angular contribution to these matrix elements is straightforward; we use the formulae of Ref.~\cite{zimmerman:1979} for $\mu_z$, generalizing for $\mu_x$ and $\mu_y$.

\subsection{Floquet Hamiltonian}
\label{se:floqhamil}

As written, Eq. \ref{eq:basic_hamiltonian} describes a time-dependent problem.  However, as the dressing field and thus the Hamiltonian is periodic ($H(t) = H(t+T)$,  with $T = 2\pi/\omega_d$), Floquet's theorem \cite{eastham:1973} may be used to write the general solution of the time-dependent Schr\"{o}dinger equation 
($ i\hbar \: \partial_t \ket{\psi(t)} = H(t) \ket{\psi(t)}$)
using the expansion \cite{shirley:1965}:
\begin{equation}
\ket{\psi(t)} = \sum_{k} c_k {\ket {\phi_k(t)}} e^{-i E_k t/\hbar}
\end{equation}
where the $c_k$'s are constant coefficients set by the initial conditions,
the $\ket{\phi_k(t)}$'s are periodic ($\ket{\phi_k(t+T)} = \ket{\phi_k(t)}$), and the $E_k$'s are known as the {\em quasi-energies} of the dressed system. In what follows, we shall simply refer to these as the dressed energies.   Since the $\ket{\phi_k(t)}$'s are periodic, they may be expanded in a Fourier series:
\begin{equation}
\ket{\phi_k(t)} = \sum_{q = 0, \pm 1, \pm 2, \dots} \ket{\tilde{\phi}_k(q)} e^{i q \omega_d t}.
\end{equation}
To describe the $\ket{\phi_k(t)}$'s, we use tensor product notation, writing the basis vectors as
$\ket{n\ell m} \otimes \ket{q} \equiv \ket{n\ell m} e^{i q \omega_d t}$.

The periodic Hamiltonian may also be expanded in a Fourier series:
\begin{equation}
H(t) = \sum_{p = 0, \pm 1, \pm 2, \dots} \tilde{H}(p) e^{i p \omega_d t}.
\end{equation}
As shown by Shirley \cite{shirley:1965}, the determination of the $\ket{\phi_k(t)}$'s and $E_k$'s that satisfy the time-dependent Schr\"{o}dinger equation amounts to an eigenvalue problem
\begin{equation}
H_F \ket{\phi_k(t)} = E_k \ket{\phi_k(t)},
\end{equation}
in which the {\em Floquet Hamiltonian} $H_F$ can be written as
\begin{equation}
H_F = 
\identity \otimes \sum_{q} \hbar \omega_d q \ket{q}\bra{q}
+\sum_{q,p} \tilde{H}(p) \otimes \ket{q+p}\bra{q} .
\label{eq:h_f}
\end{equation}
The Hamiltonian of Eq.~\ref{eq:basic_hamiltonian} allows $H_F$ to be written in a more specific form:
\begin{eqnarray}
H_F = && (H_0 - \vect{\mu} \cdot \vect{F_{\rm dc}}) \otimes \identity 
+ \identity \otimes \sum_{q} \hbar \omega_d q \ket{q}\bra{q} \nonumber \\
&& - \frac{1}{2} \sum_q \vect{\mu} \cdot \otimes \left\{ \vect{F_{ac}} \ket{q+1}\bra{q} + \vect{F_{ac}}^* \ket{q-1}\bra{q} \right\}. \nonumber \\
\label{eq:fullfloq}
\end{eqnarray}
The matrix elements of $H_F$ in the $\ket{n\ell m} \otimes \ket{q}$ basis are {\em time-independent}.  Since we must in principle include all possible Fourier components, this matrix is of infinite dimension.  In practice, $H_F$ can be approximated as a finite matrix with a limited number of ``sidebands'' \cite{vandewater:1990} and diagonalized using standard numerical methods \cite{golub:2012}.  The replacement of an infinite matrix representation of $H_F$ by a finite one is a controlled approximation --- the size of the basis set can be varied and the convergence of the eigenvalues and eigenvectors checked.  In what follows we describe this procedure as a {\em non-perturbative} Floquet calculation, to distinguish it from the perturbative approach discussed in the next section.

To characterize the basis sets used for dressed circular state calculations, we use three parameters: $\delta n$, $\delta m$, and $\delta q$.   Writing the specific circular state of interest (either $e$ or $g$) as $\ket{n_c\ell_cm_c}$, with $\ell_c = n_c-1$ and $m_c = \ell_c$,  the basis set used includes all states 
$\ket{n'\ell' m'} \otimes \ket{q'}$ with {\em valid} quantum numbers 
($\ell' < n'$ and $|m'| \le \ell'$) 
in the range:
$n' = n_c-\delta n, \dots, n_c+ \delta n$,
$m' = m_c - \delta m, \dots m_c+\delta m$ and
$q' = -\delta q, ..., \delta q$ (all ranges are in steps of one).

When there is only a dc or linearly polarized ac electric field present, or they are both present and parallel, the quantization axis for the angular momenta is chosen to be parallel to the field.  In these situations, the Hamiltonian matrix separates into diagonal blocks for different $m$, and the diagonalization effort is significantly reduced.  However in some of the calculations that follow we must consider that $\vect{F_{ac}}$ and $\vect{F_{dc}}$ are not parallel, or in general that the dressing field is not linearly polarized \cite{jones:2013}. In these cases, the operator $L_z$ no longer commutes with the Hamiltonian, and thus the couplings between states of different $m$ must be considered.

The parameters $\delta n$, $\delta q$ and $\delta m$ are increased until it is observed that the dressed energy --- obtained by diagonalization in the corresponding basis --- is changing systematically with basis set size and shows rapid convergence to a fixed value.  Unless otherwise stated, we have used $\delta n = 5$, $\delta q = 4$ and $\delta m = 1$.
All numerical results have been checked for basis set sensitivity.  We check that an increase in any one of $\delta n$, $\delta q$ and $\delta m$ from the canonical values changes the quoted result by less than the precision implied by the number of significant figures.

Similar considerations apply to the basis sets for calculations involving the low-$\ell$ momentum states of Rb.  A formal description is slightly complicated by the non-zero quantum defects.  For a state of interest 
$\ket{n \ell j m_j}$, we define $n^{*}$ using the Rydberg energy: $E_{n\ell j} = (-R_{\rm Rb})/{n^*}^2$ (where $R_{\rm Rb}$ is the Rydberg constant adjusted for the reduced mass of $e^-$-Rb$^+$ system).
The basis sets used are characterized by $\delta n$, $\delta m_j$, and $\delta q$.  They include all states $\ket{n'\ell 'j'{m_j}'} \otimes \ket{q'}$ with the zero-field energies of the undressed portion $\ket{n'\ell 'j'{m_j}'}$  satisfying
\begin{equation}
-\frac{R_{\rm Rb}}{(n^*-\delta n)^2} \le E_{n'\ell 'j'} \le -\frac{R_{\rm Rb}}{(n^*+\delta n)^2},
\end{equation}
together with ${m_j}'$ ranging over $m_j-\delta m_j, ..., m_j+\delta m_j$ and $q'$ ranging over $-\delta q, ... , \delta q$, both in steps of 1.  All calculations are done with $\delta n = 4.5$, $\delta m_j = 1$, and $\delta q = 2$.   The basis size sensitivity of all results are checked in a similar manner to the circular states.

The hyperfine structure of the Rb Rydberg states is ignored.  Previous experimental work \cite{jones:2013} was done with the $^{87}$Rb isotope, but all of the results obtained here are valid for both abundant isotopes.

\subsection{Perturbative approach}

Although numerical diagonalization of the Floquet Hamiltonian is straightforward, more insight can often be obtained by examination of the dominant terms in a perturbation expansion of the dressed energies in terms of the field strengths.  
For this purpose we partition Eq.~\ref{eq:fullfloq} 
into a zero-field Hamiltonian:
\begin{equation}
H_{F,0} = H_0 \otimes \identity 
+ \identity \otimes \sum_{q} \hbar \omega_d q \ket{q}\bra{q}
\end{equation}
and a perturbation term, including both ac and dc fields:
\begin{eqnarray}
V = && - \frac{1}{2} \sum_q \vect{\mu} \cdot \otimes \left\{ \vect{F_{ac}} \ket{q+1}\bra{q} + \vect{F_{ac}}^* \ket{q-1}\bra{q} \right\} \nonumber \\
&& - \vect{\mu} \cdot \vect{F_{\rm dc}} \otimes \identity 
\end{eqnarray}
such that $H_F = H_{F,0} + V$.  
With basis states of the form $\ket{n \ell m}\otimes\ket{q}$, the matrix elements of both $V$ and $H_{F,0}$ do not depend on time and thus time-independent perturbation theory may be used to estimate the 
$\ket{\phi_k(t)}$ and $E_k$'s.  

We will initially consider a linearly polarized ac field, parallel to the dc field:
$\vect{F_{dc}} = F_{dc} \vect{\hat{z}}$, 
$\vect{F_{ac}} = F_{ac} \vect{\hat{z}}$, with $F_{ac}$ real.
Under these conditions non-degenerate Rayleigh-Schr\"{o}dinger perturbation theory may be applied to determine the dressed energies.

\sloppy
Both the dc and ac Stark shifts are non-vanishing at second order in $V$; to examine the influence of ac fields on the dc Stark effect, we must go to fourth order.  The fourth order shift of the $k$th state is (see for example \cite{huby:1961}):
\begin{eqnarray}
\Delta E_k^{(4)} = && \sum_{u,v,w} 
\frac{
\bra{k} V \ket{u} 
\bra{u} V \ket{v} 
\bra{v} V \ket{w}
\bra{w} V \ket{k}
}{
\left(E_k^{(0)}-E_u^{(0)}\right)
\left(E_k^{(0)}-E_v^{(0)}\right)
\left(E_k^{(0)}-E_w^{(0)}\right)
}
\nonumber\\ &&
- \sum_{u,v}
\frac{
|\! \bra{k} V \ket{u} \! |^2 \:
|\! \bra{k} V \ket{v} \! |^2 
}{
\left(E_k^{(0)}-E_u^{(0)}\right)^2
\left(E_k^{(0)}-E_v^{(0)}\right)
} \label{eq:fourthorder}
\end{eqnarray}
where the summations in $u$, $v$ and $w$ are over all states other than $k$.  We have dropped terms containing the matrix elements $\bra{k} V \ket{k}$ as these couplings vanish in the normal zero dc-field Rydberg state spherical basis. The terms in this series may be grouped according to their dependence on $F_{dc}$ and $F_{ac}$:
\begin{eqnarray}
\Delta E_{k}^{(4)} = & \: C_k(4,0) \: {F_{dc}}^4 
 + C_k(2,2,\omega_d) \: {F_{dc}}^2 \left( F_{ac}/2 \right)^2  \nonumber\\
& + C_k(0,4,\omega_d) \: \left( F_{ac}/2 \right)^4 
,
\end{eqnarray}
where the $C_k(a,b,\omega_d)$'s have been introduced to represent terms in groupings by $C_k(a,b,\omega_d) (F_{dc})^a (F_{ac}/2)^b$.   For $C_k$'s with no dressing frequency dependence ($b=0$), $\omega_d$ is dropped from the argument list.  In this notation the second order dc Stark shift is $\Delta E_k = C_k(2,0) {F_{dc}}^2$, whereas the second order ac Stark shift is $\Delta E_k = C_k(0,2,\omega_d) (F_{ac}/2)^2$.  

For polarizability nulling, we are interested in the terms that are quadratic in $F_{dc}$:
\begin{equation}
\Delta E_k \approx {F_{dc}}^2 
\: [ \;
C_k(2,0)+
\sum_{j=2,4 \dots} C_k(2,j, \omega_d) (F_{ac} / 2)^j
].
\end{equation}
If higher order terms ($j \ge 4)$ can be neglected, polarizability nulling can be obtained when $C_k(2,0)+C_k(2,2,\omega_d) (F_{\rm ac}/2)^2 = 0$.  This requires working at a dressing frequency such that $C_k(2,0)$ and $C_k(2,2,\omega_d)$ are of opposite sign.  When there are two states of interest (e.g.~$e$ and $g$), the dc differential polarizability is characterized by $C_{e,g}(2,0) \equiv C_{e}(2,0)-C_{g}(2,0)$, and to obtain nulling it is necessary that $C_{e,g}(2,0)$ and $C_{e,g}(2,2,\omega_d)$ be of opposite signs.

The dipole matrix elements which contribute to the $C_k$'s are fixed, whereas the denominators in the perturbation expansion can be varied by tuning the dressing frequency.  To determine which terms in the perturbation expansion of Eq.~\ref{eq:fourthorder} are dominant, we examine when their denominators show small energy differences (i.e.~are near-resonant).

\section{Results}

\subsection{Circular Rydberg states with collinear dc and dressing fields}

For concreteness --- as with Ref.'s \cite{hyafil:2004} and \cite{mozley:2005} --- we study the electric field sensitivity of transitions between the $n=50$ and $n=51$ circular Rydberg states ($e$ and $g$ hereafter).  In the absence of a dressing field, these states exhibit a differential dc polarizability characterized by $C_{e,g}(2,0) \approx -25.44\unit{Hz/(V/m)^2}$. (Machine readable definitions of most calculated quantities and parameters in this manuscript are available in Ref.~\cite{numbers_for_dressed_nulling_theory:2015}.)

Consider the calculation of $C_c(2,2,\omega_d)$ using Eq.~\ref{eq:fourthorder} (where $c$ stands for a generic circular state $\ket{n,\ell=n-1,m=n-1}$).  Due to the energy degeneracy of high-$\ell$ states with the same $n$,  all three energy denominators in the first summation term of Eq.~\ref{eq:fourthorder} are resonant when 
$\ket{w} = \ket{n+1, n , n-1} \otimes \ket{q=-1}$
$\ket{u} = \ket{n+1, n-1 , n-1} \otimes \ket{q=-1}$ and
$\ket{v} = \ket{w}$.
For this term the values of $E_c^{(0)} - E_w^{(0)}$, $E_c^{(0)} - E_u^{(0)}$, and $E_c^{(0)}-E_v^{(0)}$ are the same ($\equiv \delta$) because states of the same $n$ but different $\ell$ are energy degenerate --- a situation that does not occur for the lower-$\ell$ states of non-hydrogenic systems.  

\begin{figure}
\includegraphics{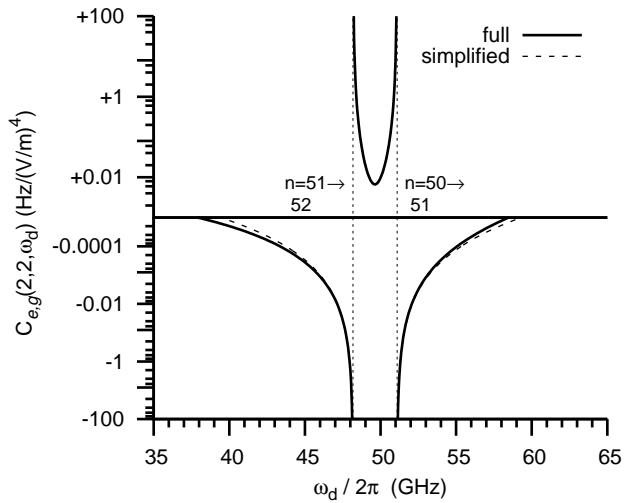}
\caption{\label{fg:circular22}
Coefficient for the nulling of the differential polarizability between
the two circular states $n=50$, $\ell = 49$, $m = 49$ and $n=51$, $\ell = 50$, $m = 50$.  Note the discontinuity in the vertical axis.
The positive $C_{e,g}(2,2,\omega_d)$ at dressing frequencies between the two resonances ($n=51 \rightarrow 52$ and $n=50 \rightarrow 51$) allows the differential dc polarizability to be nulled by a dressing field with $F_{ac}$ satisfying $C_{e,g}(2,0) + C_{e,g}(2,2, \omega_d) (F_{ac}/2)^2 = 0$.
``Simplified'' refers to the use of Eq.~\ref{eq:simplified_c_2_2}.
}
\end{figure}

This single {\em triply} resonant term with a $1/\delta^3$ dependence will dominate over other terms in the series ($1/\delta^2$, $1/\delta$) as the resonance condition is approached.  Note that the second set of summations in Eq.~\ref{eq:fourthorder} cannot contribute any $1/\delta^3$ terms to $C_c(2,2,\omega_d)$. With $\omega_d$ approximately resonant with the $n \rightarrow n+1$ transition frequency $\omega_{e,g}$ we have (in atomic units):
\begin{equation}
C_c(2,2,\omega_{e,g}+\delta) \approx 
\frac{
  |\!\bra{u_{\rm b}} \mu_z \ket{w_{\rm b}}\!|^2 \:
  |\!\bra{w_{\rm b}} \mu_z \ket{c_{\rm b}}\!|^2
}{
  \delta^3
}
\label{eq:simplified_c_2_2}
\end{equation} 
where the matrix elements are between the ``bare'' atomic states i.e.~$\ket{w_{\rm b}} = \ket{n+1, n, n-1}$, etc.
In Fig.~\ref{fg:circular22} we show $C_{e,g}(2,2,\omega_d)$ as a function of frequency, calculated using this simplified form (with two terms, each corresponding to either $e$ or $g$) and the general series summation of Eq.~\ref{eq:fourthorder}.  As can be seen, these single terms dominate the behaviour near resonance, greatly simplifying our understanding of this problem.  

\begin{figure}
\includegraphics{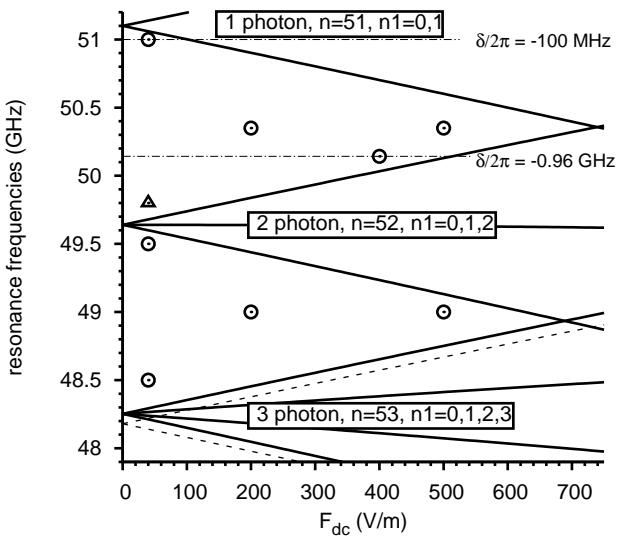}
\caption{\label{fg:resonance_soup}
Resonance frequencies of transitions from the $g$-state (solid lines) and $e$-state (two dashed lines) as function of dc electric field, over the approximate frequency range where $C_{e,g}(2,2,\omega_d)$ is positive (see Fig.~\ref{fg:circular22}).  The two dashed lines correspond to $e \rightarrow n=52, n_1=0,1$ (both $m=51$) transitions.   Two specific frequencies (dot-dash) discussed in the text are labelled by their detunings $\delta/2\pi$ from the $n=50 \rightarrow n=51$ transition frequency.   As discussed in Section \ref{se:noncollinear}, the role of orthogonal field fluctuations are investigated at the points marked with {\Large $\circ$}.   It was not possible to find a dipole null at the dc field and dressing frequency marked with $\bigtriangleup$.
}
\end{figure}

The triply resonant condition found for circular states  (Eq.~\ref{eq:simplified_c_2_2}) is exceptional --- for low-$\ell$ states there are only doubly resonant terms in the perturbation expansion and these do not change sign as we go through a resonance.  Normally, the dressed contribution to the polarizability has the {\em same sign} on both sides of a resonance -- we are ``stuck'' with whatever the resonance gives us.  As discussed later this is illustrated by the low-$\ell$ Rydberg states of Rb \cite{jones:2013}.  In contrast, we can make a circular Rydberg state either more or less polarizable by sitting on either side of the $n \rightarrow n+1$ transition.  Considering that $C_c(2,0)$ is always negative, the {\em absolute} polarizability of a circular Rydberg state with principal quantum number $n$ can be nulled using a 
$C_c(2,2,\omega_d)$ that is positive i.e.~by setting $\omega_d$ slightly greater in frequency than the $n \rightarrow n+1$ transition frequency $\omega_{e,g}$, so that $\delta > 0$.

To null the polarizability difference between two circular Rydberg states $n$ and $n+1$, it is possible to choose a dressing frequency that increases the Stark shift of the lower-$n$ state, while decreasing the Stark shift of the upper-$n$ state by working at a frequency intermediate between the $n+1 \rightarrow n+2$ (lower) and $n \rightarrow n+1$ (upper) transition frequencies.

Given the flexibility in the choice of the dressing frequency, how should it be chosen?  In Ref.~\cite{jones:2013}, it was shown that a dressing frequency could be chosen to obtain dipole nulling {\em and} set the differential ac Stark shift between the states equal to zero (known as a ``magic wavelength'' in the optical domain \cite{ye:2008}).  This has the clear advantage that spatial inhomogeneities in the dressing field over a sample will have a reduced effect on the transition frequency.  (The inhomogeneities will however lead to an reduction of the efficacy of the nulling.)

Unfortunately, for the $e \rightarrow g$ circular-circular transition considered here, the $C_{e,g}(0,2)$ coefficient that determines the differential ac Stark shift is always positive and shows no zero in the frequency range of positive $C_{e,g}(2,2)$ suitable for nulling (see Fig.~\ref{fg:circular22}). (The upper $e$ state is blue-shifted and the lower $g$ state is red-shifted.)

However, it is possible to minimize the ac differential Stark shift:  
as $C_{e,g}(0,2)$ scales like $1/\delta^2$, whereas $C_{e,g}(2,2)$ scales like $1/\delta^3$, a lower Stark shift under polarizability nulling conditions can be obtained with smaller $\delta$.

As $\delta$ is decreased for a lower ac Stark shift, one must consider that sources of the dressing field will not be perfect, with spectral impurity of increasing magnitude closer to the carrier frequency (see for example Ref.~\cite{keysight:5990-7529EN}).  This noise can drive the resonant transition \cite{jones:2013} --- an undesirable situation.  However it is also true that to obtain nulling, a weaker dressing field amplitude is required as the dressing frequency approaches the resonance.  An optimal compromise between these competing effects will depend on the spectral noise properties of the available dressing source.  Somewhat arbitrarily, we choose to study $\delta/2\pi=-100\unit{MHz}$, corresponding to an offset from the zero dc-field resonance beyond which the phase noise density of typical microwave sources is white \cite{keysight:5990-7529EN}.

Figure \ref{fg:resonance_soup} shows some relevant resonances as a function of dc field.  The choice of $\delta/2\pi = -100 \unit{MHz}$ avoids multi-photon resonances, such as those that occur at $\approx 49.65\unit{GHz}$.

With a detuning of $100\unit{MHz}$ to the red of the $n=50 \rightarrow 51$ transition ($\delta / 2\pi = -100\unit{MHz}$), summation of the relevant terms of Eq.~\ref{eq:fourthorder} gives $C_{e,g}(2,2,\omega_{e,g}+\delta) \approx 9.968 \unit{Hz/(V/m)^4}$  (application of the simple Eq.~\ref{eq:simplified_c_2_2} gives $C_{e,g}(2,2,\omega_{e,g}+\delta) \approx -C_g(2,2,\omega_{e,g}+\delta) \approx 9.90 \unit{Hz/(V/m)^4}$).  With $C_{e,g}(2,0) \approx -25.44\unit{Hz/(V/m)^2}$ this predicts that $F_{ac} \approx 3.195 \unit{V/m}$ will eliminate the 2nd order Stark effect --- so-called {\em polarizability nulling}.

\begin{figure}
\includegraphics{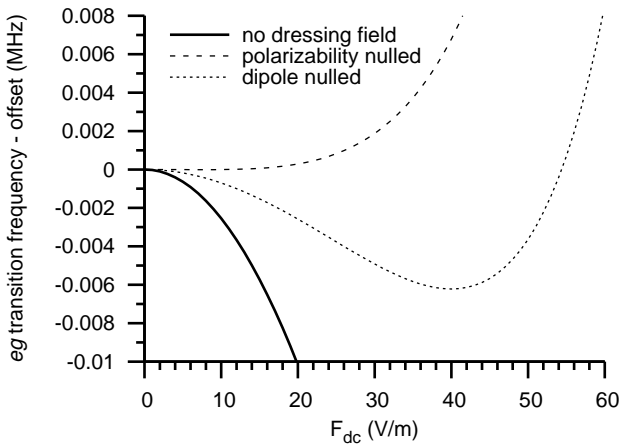}
\caption{\label{fg:circ_weg_vs_fdc_parallel} Influence of dressing fields on the dc electric field dependence of the 
$\ket{n=50,\ell=49,m=49} \rightarrow
\ket{n=51,\ell=50,m=50} $ ($e \rightarrow g$)
circular to circular state transition with parallel dc and dressing fields.  These are calculated by diagonalization of the Floquet Hamiltonian for varying $F_{dc}$.
For comparison, the transition frequency with no dressing fields is shown.  In both the polarizability- and dipole-nulled cases the dressing field frequency
is $100\unit{MHz}$ to the red of the zero ac/dc field $eg$ transition.  For polarizability nulling, $F_{ac} \approx 3.195 \unit{V/m}$; and for dipole nulling,
$F_{ac} \approx 2.718 \unit{V/m}$.  The ac Stark shifts in zero dc electric field of the polarizability-nulled and dipole-nulled cases are $\approx 0.273\unit{MHz}$ and $\approx 0.198\unit{MHz}$ respectively.
}
\end{figure}

Polarizability nulling with these parameters can be verified by diagonalization of the Floquet matrix as a function of applied dc field i.e.~a non-perturbative calculation.  Figure \ref{fg:circ_weg_vs_fdc_parallel} shows the $eg$ energy difference as a function of dc electric field both with and without the dressing field.  The dipole nulling case shown in the figure will be discussed shortly.

It is also possible to null the absolute polarizability of either the $g$ or $e$ state (but not both simultaneously).  More specifically, considering the $g$ state, but with a {\em blue} detuning of $\omega_d/2\pi$ from $w_{e,g}/2\pi$ of $\delta/2\pi = +100 \unit{MHz}$, we find that summation of the relevant terms of Eq.~\ref{eq:fourthorder} gives 
$C_g(2,2,\omega_d) \approx 9.83 \unit{Hz/(V/m)^4}$.  
With an absolute polarizability corresponding to  $C_g(2,0) \approx -203.24 \unit{Hz/(V/m)^2}$, a field strength of $F_{ac} \approx 9.09 \unit{V/m}$ is predicted for polarizability nulling.  A Floquet diagonalization at this $F_{ac}$ shows that quadratic variation of the dressed $g$ state energy with dc field is reduced to $\approx -17.3\unit{Hz/(V/m)^2}$.  Adjusting $F_{ac}$ to $9.54\unit{V/m}$ brings the magnitude of this quadratic variation to $< 1 \unit{Hz/(V/m)^2}$.  

The preceeding discussion demonstrates that it is possible --- in principle --- to completely null either the absolute polarizability of a circular Rydberg state, or the differential polarizability between two Rydberg states.  However, we have assumed that the dc and ac fields point in the same direction.  In practical work with circular Rydberg states \cite{raimond:2001}, small uncontrolled electric fields with random directions make it necessary to stabilize circular Rydberg states against mixing with lower $\ell$ and $m$ states \cite{gross:1986}.  For this purpose, a small dc electric field is used to remove the energy degeneracy.  For example, the experimental work of Ref.~\cite{brune:1999} used a stabilizing dc bias field of $36\unit{V/m}$.

To understand the practical aspects relating to non-parallel fields we will first consider how {\em dipole nulling} can be obtained in small dc electric fields parallel to the dressing field (viewing this as a perturbation from the polarizability nulling situation), and then investigate the influence of 1) fluctuations in electric field transverse to the deliberately applied electric field, and 2) the consequences of imperfect alignment of the dc bias and dressing fields.

Dipole nulling at small non-zero dc fields can be understood using a simple theory.  Terms with coefficients $C_k(4,0)$, and $C_k(4,i,\omega_d)$ (with $i=2, 4, \dots$) in the general expansion 
$\Delta E_k = \sum_{i,j} C_k(i,j) (F_{dc})^i (F_{ac}/2)^j$
describe the quartic variation of energies with dc field.  In the presence of perfect polarizability nulling, these predict the range of dc fields in which nulling is effective.  Consider that a quartic of the form $y=a_0 + a_2 x^2 + a_4 x^4$ has local extrema at $x = 0, \pm \sqrt{-a_2/2 a_4}$.  Likewise, if polarizability nulling is not perfect ($C_{e,g}(2,0) + C_{e,g}(2,2, \omega_d) (F_{ac}/2)^2 \ne 0$), the combination of the residual quadratic variation and quartic terms can lead to a local extrema (dipole null) at a non-zero dc electric field.

For  $|C_{e,g}(4,2,\omega_d) (F_{\rm ac}/2)^2| \gg |C_{e,g}(4,0)|$ (see below) the location of the dipole null can be estimated as:
\begin{equation}
F_{dc} \approx \left [
\frac{-C_{e,g}(2,0)-C_{e,g}(2,2,\omega_d)(F_{ac}/2)^2}{2 \: C_{e,g}(4,2,\omega_d) (F_{ac}/2)^2}
\right]^{1/2}
\label{eq:fdc_simple}
\end{equation}  
Near the $\omega_{e,g}$ resonance, the value of $C_{e,g}(4,2,\omega_d)$ can be estimated using similar considerations as for $C_{e,g}(2,2,\omega_d)$, again due to the energy degeneracies of high-$\ell$ states.  In fact, the
results of Eq.~\ref{eq:simplified_c_2_2} generalize to (in atomic units):
\begin{equation}
C_c(i,2,\omega_{e,g}+\delta) \approx 
\frac{
  |\!\bra{u_{\rm b}} \mu_z \ket{w_{\rm b}}\!|^i \:
  |\!\bra{w_{\rm b}} \mu_z \ket{c_{\rm b}}\!|^2
}{
  \delta^{i+1}
}
\label{eq:simplified_c_i_2}
\end{equation} 
for $i=0,2,4,6, \dots$.  We can verify that for the $n=50 \rightarrow 51$ system, with $\delta/2\pi = -100\unit{MHz}$, $F_{ac} \approx 3\unit{V/m}$
and  $C_{e,g}(4,2,\omega_d) \approx -C_c(4,2,\omega_{e,g}+\delta)$
that 
$|C_{e,g}(4,2,\omega_d) (F_{\rm ac}/2)^2| \gg |C_{e,g}(4,0)|$ and thus Eq.~\ref{eq:fdc_simple} is reasonable.
To put a dipole null at $40\unit{V/m}$ (approximately the same dc field used in Ref.~\cite{brune:1999}) with $\delta/2\pi = -100 \unit{MHz}$, Eq.~\ref{eq:fdc_simple} predicts $F_{ac}=2.798 \unit{V/m}$.  (Note that this $F_{ac}$ has a magnitude slightly smaller than that required for polarizability nulling since
$C_{e,g}(2,0) < 0$, 
$C_{e,g}(2,2,\omega_d) > 0$ 
and
$C_{e,g}(4,2,\omega_d) > 0$.) 
 A non-perturbative Floquet calculation using this value of $F_{ac}$ exhibits a null at $36.6\unit{V/m}$.

It is straightforward to improve on the estimate of Eq.~\ref{eq:fdc_simple}
by considering all terms with coefficients of the form $C_{e,g}(i,2,\omega_d)$.  Differentiating
\[
E_{e,g} = C_{e,g}(2,0) F_{dc}^2 + \sum_{i=0,2,4,\dots} C_{e,g}(i,2,\omega_d) (F_{dc})^i (F_{ac}/2)^2
\]
with respect to $F_{dc}$, setting this derivative equal to zero, and rearranging gives an expression for $F_{ac}$.  With  
$C_{e,g}(i,2,\omega_d) \approx -C_c(i,2,\omega_d)$ as given by Eq.~\ref{eq:simplified_c_i_2}, summation of the resulting arithmetico-geometric sequence gives (in atomic units):
\begin{eqnarray}
\label{eq:ageo}
F_{ac} = && 2 \frac{\{-\delta\}^{3/2} \{ -C_{e,g}(2,0) \} ^{1/2}}
{
  |\!\bra{u_{\rm b}} \mu_z \ket{w_{\rm b}}\!| \:
  |\!\bra{w_{\rm b}} \mu_z \ket{c_{\rm b}}\!|
} \times
\nonumber\\ &&
 \left[ 
1-
\left(
\frac{F_{dc} |\!\bra{u_{\rm b}} \mu_z \ket{w_{\rm b}}\!|}{\delta}
\right)^2
\right],
\end{eqnarray}
to obtain a null at $F_{dc}$.  For $F_{dc} = 40 \unit{V/m}$, this equation predicts $F_{ac} \approx 2.715\unit{V/m}$.  A non-perturbative Floquet calculation using this $F_{ac}$ gives a null at $F_{dc} = 40.1 \unit{V/m}$.  With $F_{ac} \approx 2.718\unit{V/m}$ the null is within $10^{-4}\unit{V/m}$ of $40\unit{V/m}$ --- see Fig.~\ref{fg:circ_weg_vs_fdc_parallel}.  

These perturbation theory results for both dipole and polarizability nulling are easy to apply and develop intuition.  However, it is prudent to confirm their predictions by diagonalization of the Floquet Hamiltonian, particularly in situations with larger $F_{dc}$ and $F_{ac}$, where the results of perturbation theory are less likely to be valid.
For example, the Appendix discusses the dipole nulling situation of
Ref.'s \cite{hyafil:2004} and \cite{mozley:2005}, where $F_{dc} \approx 400 \unit{V/m}$ and $F_{ac} \approx 88\unit{V/m}$ --- both much greater than for the dipole null at $F_{dc} = 40 \unit{V/m}$.  With $\delta \approx -957 \unit{MHz}$ and $F_{ac} \approx 88 \unit{V/m}$, Eq.~\ref{eq:ageo} predicts a $F_{dc} \approx 265 \unit{V/m}$. Figure \ref{fg:hm_compare}b shows a dipole null at $328 \unit{V/m}$, in only moderate agreement with Eq.~\ref{eq:ageo}.

The qualitative difference between $\omega_{e,g}$ for the larger $F_{dc}$'s in Fig.'s  \ref{fg:circ_weg_vs_fdc_parallel} and \ref{fg:hm_compare} (including the presence of the second null in Fig.~\ref{fg:hm_compare}) can be explained through examination of Fig.~\ref{fg:resonance_soup} where the two dressing frequencies are marked with horizontal lines.  The upwards trend of the nulled cases in Fig.~\ref{fg:circ_weg_vs_fdc_parallel} is due to the avoided crossing between $\ket{g} \otimes \ket{q=0}$ and $\ket{n=51,n_1=0,m=49} \otimes \ket{q=-1}$, decreasing the dressed $g$ energy, which increases the $\omega_{e,g}$ frequency.  However, at the significantly lower dressing frequency of Fig.~\ref{fg:hm_compare}, there is an avoided crossing between $\ket{g} \otimes \ket{q=0}$ and $\ket{n=52,n_1=2,m=49} \otimes \ket{q=-2}$ (a two-photon resonance) that pushes the dressed $g$ energy to higher energies, decreasing the $\omega_{e,g}$ frequency with increasing $F_{dc}$.

\subsection{Circular Rydberg states with non-collinear dc and dressing fields}
\label{se:noncollinear}

It is impossible to perfectly align the dressing and dc bias fields directions, and field fluctuations may occur in directions other than the bias field.  For these reasons it is necessary to consider the influence of non-collinear dc and dressing fields on dipole nulling.

As illustrated in  Fig.~\ref{fg:schematic}b, the circular states in zero field are energy degenerate with states of the same $n$ but lower-$m$.  In the case of cylindrical symmetry (with $z$ the axis of symmetry), states of different $m$ are not coupled and thus non-degenerate Rayleigh-Schr\"{o}dinger perturbation theory may be applied, as in the previous section.  But when the dressing and bias field are no longer orthogonal, coupled degenerate states are now involved and the non-degenerate approach is no longer applicable.  In this case we have taken the approach of direct diagonalization of the Floquet Hamiltonian.   In this situation, it is critical that for the basis set parameter $\delta m \ge 1$.  However, $\delta m=1$ is entirely adequate for examination of the second order effects to be discussed here.  

For concreteness, we consider the ac field to be aligned in the $z$ direction and the dc field to deviate from this direction.  In particular, the scenario 
of interest is that $\vect{F}_{\rm dc}$ and $\vect{F}_{\rm ac}$  are initially perfectly aligned, and that the dipole-nulling situation has been achieved i.e.~there is no first order dependence on deviations of the magnitude of the dc field.  But in addition to small fluctuations of the field in this direction, we must consider field fluctations in the orthogonal direction: 
$\vect{F_{dc}} = (F_{dc,\parallel}+\Delta F_{dc,\parallel}) \hat{\vect{z}} + \Delta F_{dc,\perp} \hat{\vect{x}}$  where $F_{dc,\parallel}$ is the dc bias field discussed in the previous section.  Considering the case of Fig.~\ref{fg:circ_weg_vs_fdc_parallel}, with $F_{dc,\parallel} = 40 \unit{V/m}$, we find that to 2nd order, deviations in the transition frequency are given by 
$\Delta \omega_{e,g}/2\pi \approx k_{\perp} (\Delta F_{dc,\perp})^2 + k_{\parallel} (\Delta F_{dc,\parallel})^2$ (symmetry precludes the cross terms) with $k_{\parallel}$ and $k_{\perp}$ shown in the first row of Table \ref{tb:null_ortho}.  The magnitude of these $k$'s may be compared to
the polarizability difference in zero dc field with no dressing fields present $C_{e,g}(2,0) \approx -25.44\unit{Hz/(V/m)^2}$.  The relatively strong dependence of $\omega_{e,g}$ on the transverse field fluctuations (i.e.~large $k_{\perp}$) possibly limits the effectiveness of dipole nulling.  For example, near a polycrystalline metal surface, patch fields show comparable fluctuations in the directions both normal and orthogonal to the surface \cite{carter:2011}.  However, excess technical noise contributions are often larger in a specific direction \cite{schindler:2015}.

In the absence of a general theory, we have performed a survey of $k_{\parallel}$ and $k_{\perp}$ for dressing frequencies where $C_{e,g}(2,2,\omega_d)$ is positive (see Fig.~\ref{fg:circular22}).  For a given $\omega_d$ we have computed the required ac field strength to put dipole nulls at specified $F_{dc}$'s.  The results are summarized in Table \ref{tb:null_ortho} (which includes the scenario of Ref.'s~\cite{hyafil:2004,mozley:2005}), indicating that there is a slight trade-off between the variations observed in the two directions.
Determining whether nulling provides an advantage over the normal situation with a bias field --- with a linear dependence of the transition energy for fluctuations in one direction --- depends on the specific magnitudes of the fluctuating fields one is trying to mitigate against.

\begin{table}[b]
\caption{
\label{tb:null_ortho}
The sensitivities of the dressed circular-circular $g \rightarrow e$ transition frequencies to variations in the dc field in directions both parallel ($k_{\parallel}$) and perpendicular ($k_{\perp}$) to the dc bias field and dressing field.  For a given dressing frequency $\omega_d/2\pi$, the dressing field amplitude $F_{ac}$ required to put a dipole null at $F_{dc,\parallel}$ has been computed ($F_{dc,\parallel}$ and $\omega_d/2\pi$ are labelled as points Fig.~\ref{fg:resonance_soup}).  The first two lines of this table correspond to situations discussed extensively in the main text ($\delta/2\pi = -100\unit{MHz}$) and the Appendix (a comparison with Ref.'s~\cite{hyafil:2004,mozley:2005}).  Except where noted, the values of $\omega_d/2\pi$ and $F_{dc,\parallel}$ are exact, whereas the others are displayed to a precision consistent with their estimated uncertainty.
}
\begin{ruledtabular}
\begin{tabular}{ddddd}
\multicolumn{1}{c}{$\omega_d/2\pi$} &
\multicolumn{1}{c}{$F_{dc,\parallel}$} &
\multicolumn{1}{c}{$F_{ac}$} &
\multicolumn{1}{c}{$k_{\parallel}$} &
\multicolumn{1}{c}{$k_{\perp}$} 
\\
\multicolumn{1}{c}{Hz} &
\multicolumn{2}{c}{($\unit{V}/\unit{m}$)} &
\multicolumn{2}{c}{$\unit{Hz/(V/m)^2}$}  
\\ \colrule
\approx 51.00 & 40.0 & 2.7 & 18.3 & 5820 \\
\approx 50.14 & 400.0 & 88.0 & -15.9 & 6700 \\
50.35 & 200.0 & 62.1 & 6.8 & 16900 \\
50.35 & 500.0 & 37.2 & 73.1 & 969 \\
49.50 & 40.0 & 70.4 & 5.7 & 368000 \\
49.00 & 200.0 & 69.7 & 5.2 & 15000 \\
49.00 & 500.0 & 49.3 & 48.4 & 1170 \\
48.50 & 40.0 & 17.7 & 1.5 & 47000 \\
\end{tabular}
\end{ruledtabular}
\end{table}

In addition to random field fluctuations, we must also consider systematic misalignment between the bias and dressing field directions.  
Small angular displacements of these two fields above a threshold value of between 3 and 4 degrees can completely eliminate the dipole nulls at $F_{dc} \approx 328 \unit{V/m}$ and $\approx 400\unit{V/m}$, as shown in Fig.~\ref{fg:hm_compare} (curve b) of the Appendix (the inflection point between these two extrema vanishes).  This sensitivity to an experimental imperfection is of concern for implementation.  On the other hand, the dipole null at $F_{dc} \approx 40 \unit{V/m}$ is Fig.~\ref{fg:circ_weg_vs_fdc_parallel} is more robust: the null remains for angular field separations up to at least 10 degrees.

\subsection{Rb Rydberg $s$-states with non-collinear dc and dressing fields}

In light of the high sensitivity of the dipole-nulled circular-circular transitions to transverse field fluctuations --- as discussed in the previous section --- we now consider the transverse field sensitivity of the dipole-nulled Rb, $49s_{1/2} \rightarrow 48s_{1/2}$ two-photon transition studied previously \cite{jones:2013}.

Without any dressing field applied, the $49s_{1/2}-48s_{1/2}$ transition frequency has a quadratic sensitivity to dc field characterized by $C_{49s,48s}(2,0) \approx -294 \unit{Hz/(V/m)^2}$.
Selection of the dressing field frequency and amplitude required for polarizability nulling is similar to that presented for the circular case.  A plot of $C_{49s,48s}(2,2,\omega_d)$ allows identification of dressing frequency ranges where nulling is possible ($C_{49s,48s}(2,2,\omega_d) > 0$, so that $C_{49s,48s}(2,0) + C_{49s,48s}(2,2, \omega_d) (F_{ac}/2)^2 = 0$ may be satisfied).  Unlike in the circular case, it is possible within at least one of these frequency ranges to satisfy $C_{49s,48s}(0,2,\omega_d) = 0$.  This eliminates the 2nd order differential ac Stark shift and makes the transition less sensitive to variations in ac field strength (e.g.~due to spatial inhomogeneities).

Plots of $C_{49s,48s}(0,2,\omega_d)$ and $C_{49s,48s}(2,2,\omega_d)$ vs.~$\omega_d$ are similar to the plots presented in Fig.~1 of Ref.~\cite{jones:2013} for $\Delta \alpha(\omega_d)$ and $\Delta \beta(\omega_d)$, with the correspondence: 
$C_{49s,48s}(0,2,\omega_d) \approx - \Delta \alpha(\omega_d)$
and
$C_{49s,48s}(2,2,\omega_d) \approx 2 \Delta \beta(\omega_d) / F_{dc,0}$ 
where $F_{dc,0} = 100 \unit{V/m}$.  In Ref.~\cite{jones:2013}, $\Delta \beta(\omega_d)$ is computed using third-order perturbation theory.  The ``unperturbed'' eigenstates are those obtained by diagonalization of the Stark Hamiltonian in a non-zero dc electric field ($F_{dc,0} = 100 \unit{V/m}$).  Deviations from this dc field together with the ac field constitute the perturbation. Likewise $\Delta \alpha (\omega_d)$ was computed in a dc field of $100 \unit{V/m}$ and thus is approximately equal but not identical to $C_{49s,48s}(0,2,\omega_d)$.

For consistency, we choose $\omega_d/2\pi = 38.465 \unit{GHz}$ as in Ref.~\cite{jones:2013}, which is slightly displaced from the $C_{49s,48s}(0,2,\omega_d)=0$ condition, but gives roughly zero differential ac Stark shift at a dc field of $\approx 100 \unit{V/m}$.  With this $\omega_d$, we compute $C_{49s,48s}(2,2,\omega_d) \approx 7.18 \unit{Hz/(V/m)^4}$, and thus $F_{ac} \approx 12.79 \unit{V/m}$ should lead to polarizability nulling (using $C_{49s,48s}(2,0)$ mentioned previously).  A non-perturbative Floquet calculation with a basis set suitable for the Rb states (see discussion in Section \ref{se:floqhamil}) confirms that the dc polarizability is reduced to $\approx 10 \unit{Hz/(V/m)^2}$ under these conditions.  The value of $F_{ac}$ can be tweaked to completely eliminate this polarizability, or to place a dipole null at a specific field, as in the circular-circular case.  With $F_{ac} \approx 13.01 \unit{V/m}$, a dipole-null can be placed at $F_{dc} = 110 \unit{V/m}$.

The effect of non-parallel $\vect{F_{dc}}$ and $\vect{F_{ac}}$ can be determined using diagonalization of the Floquet Hamiltonian.  The effect of transverse field fluctations can be characterized in a similar way to the circular-circular case: for a dipole null at $F_{dc,\parallel} = 110 \unit{V/m}$, we have 
$k_{\parallel} \approx 70 \unit{Hz/(V/m)^2}$,  $k_{\perp} \approx 34 \unit{Hz/(V/m)^2}$, 
both of which are relatively small compared to $C_{49s,48s}(2,0) \approx -294 \unit{Hz/(V/m)^2}$.

There is a reduced sensitivity to transverse fluctuations for the dipole-nulled Rb $49s_{1/2}-48s_{1/2}$ transition, as compared to the circular-circular case (i.e.~lower $k_{\perp}$ for the Rb case).   This is not surprising given the isotropy of the unperturbed $s$-states compared to circular states.    We have also found a relatively small sensitivity to transverse fluctuations in the case of a dipole-nulled $s-p$ transition between triplet Rydberg states of He (manuscript in preparation).

\section{Concluding remarks}

Summarizing:
\begin{enumerate}
\item Circular Rydberg states can be polarizability nulled, in a similar manner to low-$\ell$ Rydberg states \cite{jones:2013}.  Both the absolute and the differential polarizability between consecutive $n$, $n+1$ circular Rydberg states can be nulled.  The corresponding perturbation theory description is relatively simple compared to the case of low-$\ell$ states, where numerical summation of the perturbation series is required.
\item By considering deviations from the polarizability-nulling situation, the parameters required for the dipole-nulling of circular Rydberg states \cite{hyafil:2004,mozley:2005} can be calculated in a straightforward manner.
\item Under the conditions of circular state dipole-nulling, large energy shifts are observed for small dc electric field perturbations transverse to dressing and dc bias fields (both parallel).
This may limit the usefulness of dipole nulling depending on the specifics of the electric field fluctuations.
\item Dipole-nulling of transitions between low-$\ell$ Rydberg states --- such as the $49s_{1/2}-48s_{1/2}$ transition in Rb ---  do not show the same high sensitivity to transverse fields as circular states.
\end{enumerate} 
Any planned experimental work on the reduction of electric field sensitivities of circular Rydberg states should take the third point into careful consideration.

Low-$\ell$ Rydberg states remain interesting targets for the reduction of dc electric field sensitivity using dressing fields --- without large transverse field sensitivity.  Particularly interesting is the possibility of dipole-nulling single-photon transitions between Rydberg states of relatively low-$\ell$ optically accessible Rydberg states.

\acknowledgments

We thank J.~Lambert, A.~Anderson, and W.~Cui for comments on this manuscript.  This work was supported by NSERC.

\appendix

\section{Comparison with previous dressed circular Rydberg atom calculations}

The techniques that we have used in the main body agree qualitatively but not quantitatively with previous circular Rydberg state dressing calculations \cite{hyafil:2004,mozley:2005}.  Using Ref.~\cite{mozley:2004} we have determined the approximations required for us to obtain agreement with this earlier work.  As this information is not widely available, we describe these approximations and the resulting deviations from our approach.

References \cite{hyafil:2004} and \cite{mozley:2005} show that a linearly polarized non-resonant dressing field can suppress the difference between the permanent electric dipole moments of two dressed circular states, $n=50$, $\ell=m=49$ ($g$ hereafter) and $n=51$, $\ell=m=50$ ($e$ hereafter) in a dc field of 400\unit{V/m}.  In other words, the transition frequency between these dressed states is insensitive to first-order dc electric field fluctuations (``dipole-nulled''). 

The dipole nulling condition can be obtained using different combinations of dressing frequencies and field strengths.  In Ref.'s \cite{hyafil:2004,mozley:2005} the field frequency and amplitude are characterized using the near resonant transition: $g \rightarrow i$ where $i$ is the  $n=51$, $n_1=0$, $n_2=1$, $m=49$ parabolic state (see Fig.~\ref{fg:schematic}b).  In particular, the dressing field amplitude $F_{ac}$ is chosen such that 
$F_{ac} |\bra{i}  \mu_z  \ket{g}| / h = 200 \unit{MHz}$ (where $h$ is Planck's constant).
The transition dipole moment is computed using the undressed {\em zero} dc-field states 
(i.e.~the parabolic states are not exact energy eigenstates once a dc field is applied --- see Section 52 of Ref.~\cite{bethe:1957}).  The formulae of Ref.~\cite{bethe:1957} give
$|\bra{i}  \mu_z  \ket{g}| \approx 177.6$ (in atomic units) 
and thus $F_{ac} \approx 87.99 \unit{V/m}$.

The dressing field frequency required for nulling with this $F_{ac}$ is specified in Ref.'s \cite{hyafil:2004,mozley:2005} as a detuning {\em relative} to the undressed $g \rightarrow i$ resonance frequency at $400\unit{V/m}$.  This resonance frequency is calculated using diagonalization of the Stark Hamiltonian in a finite basis of spherical states $\ket{n'\ell 'm'}$, that includes all valid states with $m'=n-1$,
$n'$ ranging over $n$, $n+1$, $...$, $n+\delta n$, and $\ell '$ ranging from  $m'$ to $n'-1$, where $n=50$.  The parameter $\delta n$ controls the size of the basis set and subsequent accuracy of the calculation (see below); following Ref.~\cite{mozley:2005} we used $\delta n =5$ ($M=6$ in their notation). The resonance frequency at $400 \unit{V/m}$ is found to be $\approx 50.702299\unit{GHz}$. With a red detuning of $0.555907\unit{GHz}$ \cite{mozley:2005}, the actual dressing frequency is $\approx 50.146392\unit{GHz}$ \cite{numbers_for_dressed_nulling_theory:2015}.

\begin{figure}
\vspace{0.2in}
\includegraphics{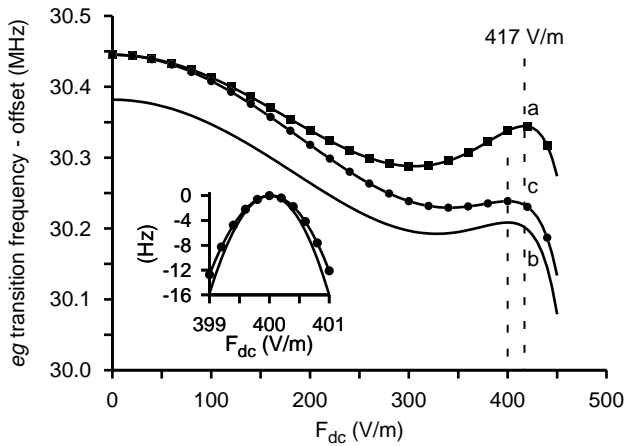}
\caption{\label{fg:hm_compare}
Transition frequencies between the dressed $g$ and $e$ states as a function of dc electric field.  The frequencies are relative to the zero dc and ac electric field transition frequency $\Delta E / h$ where $\Delta E = 0.5 \: (1/50^2-1/51^2)$ (in atomic units).  
Curve (a) uses our calculation procedure with the dressing field parameters of Ref.~\cite{mozley:2005}; curve (b) corresponds to our calculation procedure with the same dressing field amplitude as Ref.~\cite{mozley:2005}, but with a frequency chosen to place a null at 400 \unit{V/m}; curve (c) is a calculation using the HM procedure (see text) with the dressing field parameters of Ref. \cite{mozley:2005}.  
The inset compares (b) and (c) around the dipole nulled condition, relative the respective dressed transition frequencies at $F_{dc}=400\unit{V/m}$.
}
\end{figure}

This dressing field frequency and the above-mentioned amplitude can be used to compute the difference between the dressed $g$ and $e$ states as a function of dc field (see Fig.~\ref{fg:hm_compare}, curve a).   As in the main text, the Floquet method is used, with sidebands up to and including $\pm 4$ added ($\delta q=4$ in the notation of the main text; $N=4$ in the notation of Ref.~\cite{mozley:2005}).  As the two circular Rydberg states are not coupled by either the dc or ac electric field --- assuming both fields have the same linear polarization --- their Hamiltonians can be diagonalized separately.  In both cases a total of $M=6$ different $n$ manifolds are used, starting with the lowest $n$ required ($n=50$ for $g$ and $n=51$ for $e$).

When calculated in this manner, the energy difference between the $g$ and $e$ states does exhibit dipole nulling, but at a dc electric field of $\approx 417\unit{V/m}$, shifted with respect to the value of $400\unit{V/m}$ reported in Ref.'s \cite{hyafil:2004,mozley:2005}.

We find that by either changing the coupling from $200\unit{MHz}$ to $\approx 194.79 \unit{MHz}$ or by changing the dressing frequency from 
$\approx 50.146392\unit{GHz}$ to $\approx 50.142205\unit{GHz}$
(see curve b of Fig.~\ref{fg:hm_compare}), the dipole nulling condition can be obtained at 400 V/m.

Thus the procedure of our main text gives results that are inconsistent with those in Ref.'s \cite{hyafil:2004,mozley:2005} when we use the same dressing field parameters.  Examination of Ref.~\cite{mozley:2004} suggests that this difference is due to the following approximation scheme (HM hereafter): ac field couplings are computed in the parabolic basis \cite{bethe:1957}, assuming that the zero field parabolic states are the eigenkets of the {\em non-zero} dc electric field Hamiltonian.  (Expanding the zero field parabolic states in terms of the spherical states \cite{park:1960} allows computation of these couplings using Eq.'s (63.2) and (63.5) of \cite{bethe:1957} --- with the correction mentioned in Section \ref{se:techniques}.)  The energies computed using this basis after adding the dc field, but before including the ac field, are assumed to be the 2nd order perturbation energies (Eq.~52.3 of Ref.~\cite{bethe:1957}), except for the $g$, $i$ and $e$ states;  in this case the energies are obtained from a full diagonalization in a $M=6$ sized basis set (using the notation of Ref.~\cite{mozley:2005}). The essence of this approximation is that the ac field couplings do not consider that the dc field can mix states of different $n$.

With the HM approximation scheme and the parameters for the dressing field amplitude and frequency computed from Ref.~\cite{mozley:2005} (see above and \cite{numbers_for_dressed_nulling_theory:2015}) we have verified that dipole nulling occurs at a dc electric field of $400.024 \unit{V/m}$, consistent with Ref.'s \cite{hyafil:2004,mozley:2005} (see Fig.~\ref{fg:hm_compare} curve c). As expected, the HM approximation scheme agrees with the more complete calculation as the dc field goes to zero (then the ac couplings become exact).  

Figure 1b of Hyafil {\it et al.}~\cite{hyafil:2004} contains a plot with a similar axis to the inset of our Fig.~\ref{fg:hm_compare}.  When their Fig.~1b is digitized we find agreement with the HM approximation scheme to within $0.03\unit{V/m}$ and $0.25\unit{Hz}$ from $399$ to $401\unit{V/m}$.

Mozley {\it et al.}~\cite{mozley:2005} discuss the size of basis set required to reach a particular accuracy.  They found that for $\delta n \ge 5$ and $\delta q \ge 4$, a subsequent increment in either $\delta n$ or $\delta q$ did not shift the dressed energies by more than $1\unit{Hz}$.  Because it might be possible that the HM approximation scheme significantly influences the size of the basis set required, we have also checked the values of $\delta n$ and $\delta q$.

The $\delta n$ required depends on the dc electric field.  In zero dc field we find that $\delta n=5$ is adequate for both our and the HM method; but at $400\unit{V/m}$ it is necessary to go to $\delta n=7$ to meet the $1\unit{Hz}$ criteria for both methods.   However, this $1 \unit{Hz}$ convergence criteria is less stringent when applied to the energy {\em differences} as compared to the energies, which tend to shift similarly as the basis set is increased.  If we consider the dc field at which nulling is obtained, we find that curve b of the inset of Fig.~\ref{fg:hm_compare} is shifted by only $0.001\unit{V/m}$ between $\delta n=5$ and $\delta n=7$.

In summary, we conclude that good quantitative agreement is obtained with the results of Ref.'s \cite{hyafil:2004,mozley:2005} when we use the approximations described in Ref.~\cite{mozley:2004} (the HM scheme).  This comparison has been presented in the interest of reproducibility; we do not advocate use of the HM scheme as we have found it more difficult to implement and of no advantage in computational efficiency.


\begin{thebibliography}{40}\makeatletter
\providecommand \@ifxundefined [1]{ \@ifx{#1\undefined}
}\providecommand \@ifnum [1]{ \ifnum #1\expandafter \@firstoftwo
 \else \expandafter \@secondoftwo
 \fi
}\providecommand \@ifx [1]{ \ifx #1\expandafter \@firstoftwo
 \else \expandafter \@secondoftwo
 \fi
}\providecommand \natexlab [1]{#1}\providecommand \enquote  [1]{``#1''}\providecommand \bibnamefont  [1]{#1}\providecommand \bibfnamefont [1]{#1}\providecommand \citenamefont [1]{#1}\providecommand \href@noop [0]{\@secondoftwo}\providecommand \href [0]{\begingroup \@sanitize@url \@href}\providecommand \@href[1]{\@@startlink{#1}\@@href}\providecommand \@@href[1]{\endgroup#1\@@endlink}\providecommand \@sanitize@url [0]{\catcode `\\12\catcode `\$12\catcode
  `\&12\catcode `\#12\catcode `\^12\catcode `\_12\catcode `\%12\relax}\providecommand \@@startlink[1]{}\providecommand \@@endlink[0]{}\providecommand \url  [0]{\begingroup\@sanitize@url \@url }\providecommand \@url [1]{\endgroup\@href {#1}{\urlprefix }}\providecommand \urlprefix  [0]{URL }\providecommand \Eprint [0]{\href }\providecommand \doibase [0]{http://dx.doi.org/}\providecommand \selectlanguage [0]{\@gobble}\providecommand \bibinfo  [0]{\@secondoftwo}\providecommand \bibfield  [0]{\@secondoftwo}\providecommand \translation [1]{[#1]}\providecommand \BibitemOpen [0]{}\providecommand \bibitemStop [0]{}\providecommand \bibitemNoStop [0]{.\EOS\space}\providecommand \EOS [0]{\spacefactor3000\relax}\providecommand \BibitemShut  [1]{\csname bibitem#1\endcsname}\let\auto@bib@innerbib\@empty
\bibitem [{\citenamefont {Raimond}\ \emph {et~al.}(2001)\citenamefont
  {Raimond}, \citenamefont {Brune},\ and\ \citenamefont
  {Haroche}}]{raimond:2001}  \BibitemOpen
  \bibfield  {author} {\bibinfo {author} {\bibfnamefont {J.~M.}\ \bibnamefont
  {Raimond}}, \bibinfo {author} {\bibfnamefont {M.}~\bibnamefont {Brune}}, \
  and\ \bibinfo {author} {\bibfnamefont {S.}~\bibnamefont {Haroche}},\
  }\href@noop {} {\bibfield  {journal} {\bibinfo  {journal} {Rev. Mod. Phys.}\
  }\textbf {\bibinfo {volume} {73}},\ \bibinfo {pages} {565} (\bibinfo {year}
  {2001})}\BibitemShut {NoStop}\bibitem [{\citenamefont {Marcassa}\ and\ \citenamefont
  {Shaffer}(2014)}]{marcassa:2014}  \BibitemOpen
  \bibfield  {author} {\bibinfo {author} {\bibfnamefont {L.~G.}\ \bibnamefont
  {Marcassa}}\ and\ \bibinfo {author} {\bibfnamefont {J.~P.}\ \bibnamefont
  {Shaffer}},\ }in\ \href@noop {} {\emph {\bibinfo {booktitle} {Advances in
  Atomic, Molecular, and Optical Physics}}},\ Vol.~\bibinfo {volume} {63},\
  \bibinfo {editor} {edited by\ \bibinfo {editor} {\bibfnamefont
  {E.}~\bibnamefont {Arimondo}}, \bibinfo {editor} {\bibfnamefont {P.~R.}\
  \bibnamefont {Berman}}, \ and\ \bibinfo {editor} {\bibfnamefont {C.~C.}\
  \bibnamefont {Lin}}}\ (\bibinfo  {publisher} {Academic Press},\ \bibinfo
  {address} {San Diego},\ \bibinfo {year} {2014})\BibitemShut {NoStop}\bibitem [{\citenamefont {Gallagher}(1994)}]{gallagher:1994}  \BibitemOpen
  \bibfield  {author} {\bibinfo {author} {\bibfnamefont {T.~F.}\ \bibnamefont
  {Gallagher}},\ }\href@noop {} {\emph {\bibinfo {title} {Rydberg Atoms}}}\
  (\bibinfo  {publisher} {Cambridge University Press},\ \bibinfo {address}
  {Cambridge},\ \bibinfo {year} {1994})\BibitemShut {NoStop}\bibitem [{\citenamefont {Breeden}\ and\ \citenamefont
  {Metcalf}(1981)}]{breeden:1981}  \BibitemOpen
  \bibfield  {author} {\bibinfo {author} {\bibfnamefont {T.}~\bibnamefont
  {Breeden}}\ and\ \bibinfo {author} {\bibfnamefont {H.}~\bibnamefont
  {Metcalf}},\ }\href {\doibase 10.1103/PhysRevLett.47.1726} {\bibfield
  {journal} {\bibinfo  {journal} {Phys. Rev. Lett.}\ }\textbf {\bibinfo
  {volume} {47}},\ \bibinfo {pages} {1726} (\bibinfo {year}
  {1981})}\BibitemShut {NoStop}\bibitem [{\citenamefont {Procter}\ \emph {et~al.}(2003)\citenamefont
  {Procter}, \citenamefont {Yamakita}, \citenamefont {Merkt},\ and\
  \citenamefont {Softley}}]{procter:2003}  \BibitemOpen
  \bibfield  {author} {\bibinfo {author} {\bibfnamefont {S.}~\bibnamefont
  {Procter}}, \bibinfo {author} {\bibfnamefont {Y.}~\bibnamefont {Yamakita}},
  \bibinfo {author} {\bibfnamefont {F.}~\bibnamefont {Merkt}}, \ and\ \bibinfo
  {author} {\bibfnamefont {T.}~\bibnamefont {Softley}},\ }\href {\doibase
  http://dx.doi.org/10.1016/S0009-2614(03)00812-1} {\bibfield  {journal}
  {\bibinfo  {journal} {Chemical Physics Letters}\ }\textbf {\bibinfo {volume}
  {374}},\ \bibinfo {pages} {667 } (\bibinfo {year} {2003})}\BibitemShut
  {NoStop}\bibitem [{\citenamefont {Vliegen}\ \emph {et~al.}(2004)\citenamefont
  {Vliegen}, \citenamefont {W\"orner}, \citenamefont {Softley},\ and\
  \citenamefont {Merkt}}]{vliegen:2004}  \BibitemOpen
  \bibfield  {author} {\bibinfo {author} {\bibfnamefont {E.}~\bibnamefont
  {Vliegen}}, \bibinfo {author} {\bibfnamefont {H.~J.}\ \bibnamefont
  {W\"orner}}, \bibinfo {author} {\bibfnamefont {T.~P.}\ \bibnamefont
  {Softley}}, \ and\ \bibinfo {author} {\bibfnamefont {F.}~\bibnamefont
  {Merkt}},\ }\href {\doibase 10.1103/PhysRevLett.92.033005} {\bibfield
  {journal} {\bibinfo  {journal} {Phys. Rev. Lett.}\ }\textbf {\bibinfo
  {volume} {92}},\ \bibinfo {pages} {033005} (\bibinfo {year}
  {2004})}\BibitemShut {NoStop}\bibitem [{\citenamefont {Sandoghdar}\ \emph {et~al.}(1996)\citenamefont
  {Sandoghdar}, \citenamefont {Sukenik}, \citenamefont {Haroche},\ and\
  \citenamefont {Hinds}}]{sandoghdar:1996}  \BibitemOpen
  \bibfield  {author} {\bibinfo {author} {\bibfnamefont {V.}~\bibnamefont
  {Sandoghdar}}, \bibinfo {author} {\bibfnamefont {C.~I.}\ \bibnamefont
  {Sukenik}}, \bibinfo {author} {\bibfnamefont {S.}~\bibnamefont {Haroche}}, \
  and\ \bibinfo {author} {\bibfnamefont {E.~A.}\ \bibnamefont {Hinds}},\
  }\href@noop {} {\bibfield  {journal} {\bibinfo  {journal} {Phys. Rev. A}\
  }\textbf {\bibinfo {volume} {53}},\ \bibinfo {pages} {1919} (\bibinfo {year}
  {1996})}\BibitemShut {NoStop}\bibitem [{\citenamefont {Weidinger}\ \emph {et~al.}(1997)\citenamefont
  {Weidinger}, \citenamefont {Marrocco}, \citenamefont {Sang},\ and\
  \citenamefont {Walther}}]{weidinger:1997}  \BibitemOpen
  \bibfield  {author} {\bibinfo {author} {\bibfnamefont {M.}~\bibnamefont
  {Weidinger}}, \bibinfo {author} {\bibfnamefont {M.}~\bibnamefont {Marrocco}},
  \bibinfo {author} {\bibfnamefont {R.~T.}\ \bibnamefont {Sang}}, \ and\
  \bibinfo {author} {\bibfnamefont {H.}~\bibnamefont {Walther}},\ }\href@noop
  {} {\bibfield  {journal} {\bibinfo  {journal} {Opt. Commun.}\ }\textbf
  {\bibinfo {volume} {141}},\ \bibinfo {pages} {273} (\bibinfo {year}
  {1997})}\BibitemShut {NoStop}\bibitem [{\citenamefont {Hogan}\ \emph {et~al.}(2012)\citenamefont {Hogan},
  \citenamefont {Agner}, \citenamefont {Merkt}, \citenamefont {Thiele},
  \citenamefont {Filipp},\ and\ \citenamefont {Wallraff}}]{hogan:2011}  \BibitemOpen
  \bibfield  {author} {\bibinfo {author} {\bibfnamefont {S.~D.}\ \bibnamefont
  {Hogan}}, \bibinfo {author} {\bibfnamefont {J.~A.}\ \bibnamefont {Agner}},
  \bibinfo {author} {\bibfnamefont {F.}~\bibnamefont {Merkt}}, \bibinfo
  {author} {\bibfnamefont {T.}~\bibnamefont {Thiele}}, \bibinfo {author}
  {\bibfnamefont {S.}~\bibnamefont {Filipp}}, \ and\ \bibinfo {author}
  {\bibfnamefont {A.}~\bibnamefont {Wallraff}},\ }\href@noop {} {\bibfield
  {journal} {\bibinfo  {journal} {Phys. Rev. Lett.}\ }\textbf {\bibinfo
  {volume} {108}},\ \bibinfo {pages} {063004} (\bibinfo {year}
  {2012})}\BibitemShut {NoStop}\bibitem [{\citenamefont {Thiele}\ \emph {et~al.}(2014)\citenamefont {Thiele},
  \citenamefont {Filipp}, \citenamefont {Agner}, \citenamefont {Schmutz},
  \citenamefont {Deiglmayr}, \citenamefont {Stammeier}, \citenamefont
  {Allmendinger}, \citenamefont {Merkt},\ and\ \citenamefont
  {Wallraff}}]{thiele:2014}  \BibitemOpen
  \bibfield  {author} {\bibinfo {author} {\bibfnamefont {T.}~\bibnamefont
  {Thiele}}, \bibinfo {author} {\bibfnamefont {S.}~\bibnamefont {Filipp}},
  \bibinfo {author} {\bibfnamefont {J.~A.}\ \bibnamefont {Agner}}, \bibinfo
  {author} {\bibfnamefont {H.}~\bibnamefont {Schmutz}}, \bibinfo {author}
  {\bibfnamefont {J.}~\bibnamefont {Deiglmayr}}, \bibinfo {author}
  {\bibfnamefont {M.}~\bibnamefont {Stammeier}}, \bibinfo {author}
  {\bibfnamefont {P.}~\bibnamefont {Allmendinger}}, \bibinfo {author}
  {\bibfnamefont {F.}~\bibnamefont {Merkt}}, \ and\ \bibinfo {author}
  {\bibfnamefont {A.}~\bibnamefont {Wallraff}},\ }\href {\doibase
  10.1103/PhysRevA.90.013414} {\bibfield  {journal} {\bibinfo  {journal} {Phys.
  Rev. A}\ }\textbf {\bibinfo {volume} {90}},\ \bibinfo {pages} {013414}
  (\bibinfo {year} {2014})}\BibitemShut {NoStop}\bibitem [{\citenamefont {S\o{}rensen}\ \emph {et~al.}(2004)\citenamefont
  {S\o{}rensen}, \citenamefont {van~der Wal}, \citenamefont {Childress},\ and\
  \citenamefont {Lukin}}]{sorensen:2004}  \BibitemOpen
  \bibfield  {author} {\bibinfo {author} {\bibfnamefont {A.~S.}\ \bibnamefont
  {S\o{}rensen}}, \bibinfo {author} {\bibfnamefont {C.~H.}\ \bibnamefont
  {van~der Wal}}, \bibinfo {author} {\bibfnamefont {L.~I.}\ \bibnamefont
  {Childress}}, \ and\ \bibinfo {author} {\bibfnamefont {M.~D.}\ \bibnamefont
  {Lukin}},\ }\href {\doibase 10.1103/PhysRevLett.92.063601} {\bibfield
  {journal} {\bibinfo  {journal} {Phys. Rev. Lett.}\ }\textbf {\bibinfo
  {volume} {92}},\ \bibinfo {pages} {063601} (\bibinfo {year}
  {2004})}\BibitemShut {NoStop}\bibitem [{\citenamefont {Petrosyan}\ \emph {et~al.}(2009)\citenamefont
  {Petrosyan}, \citenamefont {Bensky}, \citenamefont {Kurizki}, \citenamefont
  {Mazets}, \citenamefont {Majer},\ and\ \citenamefont
  {Schmiedmayer}}]{petrosyan:2009}  \BibitemOpen
  \bibfield  {author} {\bibinfo {author} {\bibfnamefont {D.}~\bibnamefont
  {Petrosyan}}, \bibinfo {author} {\bibfnamefont {G.}~\bibnamefont {Bensky}},
  \bibinfo {author} {\bibfnamefont {G.}~\bibnamefont {Kurizki}}, \bibinfo
  {author} {\bibfnamefont {I.}~\bibnamefont {Mazets}}, \bibinfo {author}
  {\bibfnamefont {J.}~\bibnamefont {Majer}}, \ and\ \bibinfo {author}
  {\bibfnamefont {J.}~\bibnamefont {Schmiedmayer}},\ }\href@noop {} {\bibfield
  {journal} {\bibinfo  {journal} {Phys. Rev. A}\ }\textbf {\bibinfo {volume}
  {79}},\ \bibinfo {pages} {040304} (\bibinfo {year} {2009})}\BibitemShut
  {NoStop}\bibitem [{\citenamefont {Pritchard}\ \emph {et~al.}(2014)\citenamefont
  {Pritchard}, \citenamefont {Isaacs}, \citenamefont {Beck}, \citenamefont
  {McDermott},\ and\ \citenamefont {Saffman}}]{pritchard:2014}  \BibitemOpen
  \bibfield  {author} {\bibinfo {author} {\bibfnamefont {J.~D.}\ \bibnamefont
  {Pritchard}}, \bibinfo {author} {\bibfnamefont {J.~A.}\ \bibnamefont
  {Isaacs}}, \bibinfo {author} {\bibfnamefont {M.~A.}\ \bibnamefont {Beck}},
  \bibinfo {author} {\bibfnamefont {R.}~\bibnamefont {McDermott}}, \ and\
  \bibinfo {author} {\bibfnamefont {M.}~\bibnamefont {Saffman}},\ }\href
  {\doibase 10.1103/PhysRevA.89.010301} {\bibfield  {journal} {\bibinfo
  {journal} {Phys. Rev. A}\ }\textbf {\bibinfo {volume} {89}},\ \bibinfo
  {pages} {010301} (\bibinfo {year} {2014})}\BibitemShut {NoStop}\bibitem [{\citenamefont {Vion}\ \emph {et~al.}(2002)\citenamefont {Vion},
  \citenamefont {Aassime}, \citenamefont {Cottet}, \citenamefont {Joyez},
  \citenamefont {Pothier}, \citenamefont {Urbina}, \citenamefont {Esteve},\
  and\ \citenamefont {Devoret}}]{vion:2002}  \BibitemOpen
  \bibfield  {author} {\bibinfo {author} {\bibfnamefont {D.}~\bibnamefont
  {Vion}}, \bibinfo {author} {\bibfnamefont {A.}~\bibnamefont {Aassime}},
  \bibinfo {author} {\bibfnamefont {A.}~\bibnamefont {Cottet}}, \bibinfo
  {author} {\bibfnamefont {P.}~\bibnamefont {Joyez}}, \bibinfo {author}
  {\bibfnamefont {H.}~\bibnamefont {Pothier}}, \bibinfo {author} {\bibfnamefont
  {C.}~\bibnamefont {Urbina}}, \bibinfo {author} {\bibfnamefont
  {D.}~\bibnamefont {Esteve}}, \ and\ \bibinfo {author} {\bibfnamefont {M.~H.}\
  \bibnamefont {Devoret}},\ }\href@noop {} {\bibfield  {journal} {\bibinfo
  {journal} {Science}\ }\textbf {\bibinfo {volume} {296}},\ \bibinfo {pages}
  {886} (\bibinfo {year} {2002})}\BibitemShut {NoStop}\bibitem [{\citenamefont {Timoney}\ \emph {et~al.}(2011)\citenamefont
  {Timoney}, \citenamefont {Baumgart}, \citenamefont {Johanning}, \citenamefont
  {Var{\'o}n}, \citenamefont {Plenio}, \citenamefont {Retzker},\ and\
  \citenamefont {Wunderlich}}]{timoney:2011}  \BibitemOpen
  \bibfield  {author} {\bibinfo {author} {\bibfnamefont {N.}~\bibnamefont
  {Timoney}}, \bibinfo {author} {\bibfnamefont {I.}~\bibnamefont {Baumgart}},
  \bibinfo {author} {\bibfnamefont {M.}~\bibnamefont {Johanning}}, \bibinfo
  {author} {\bibfnamefont {A.}~\bibnamefont {Var{\'o}n}}, \bibinfo {author}
  {\bibfnamefont {M.}~\bibnamefont {Plenio}}, \bibinfo {author} {\bibfnamefont
  {A.}~\bibnamefont {Retzker}}, \ and\ \bibinfo {author} {\bibfnamefont
  {C.}~\bibnamefont {Wunderlich}},\ }\href@noop {} {\bibfield  {journal}
  {\bibinfo  {journal} {Nature}\ }\textbf {\bibinfo {volume} {476}},\ \bibinfo
  {pages} {185} (\bibinfo {year} {2011})}\BibitemShut {NoStop}\bibitem [{\citenamefont {Zanon-Willette}\ \emph {et~al.}(2012)\citenamefont
  {Zanon-Willette}, \citenamefont {de~Clercq},\ and\ \citenamefont
  {Arimondo}}]{zanon:2012}  \BibitemOpen
  \bibfield  {author} {\bibinfo {author} {\bibfnamefont {T.}~\bibnamefont
  {Zanon-Willette}}, \bibinfo {author} {\bibfnamefont {E.}~\bibnamefont
  {de~Clercq}}, \ and\ \bibinfo {author} {\bibfnamefont {E.}~\bibnamefont
  {Arimondo}},\ }\href@noop {} {\bibfield  {journal} {\bibinfo  {journal}
  {Phys. Rev. Lett.}\ }\textbf {\bibinfo {volume} {109}},\ \bibinfo {pages}
  {223003} (\bibinfo {year} {2012})}\BibitemShut {NoStop}\bibitem [{\citenamefont {S\'ark\'any}\ \emph {et~al.}(2014)\citenamefont
  {S\'ark\'any}, \citenamefont {Weiss}, \citenamefont {Hattermann},\ and\
  \citenamefont {Fort\'agh}}]{sarkany:2014}  \BibitemOpen
  \bibfield  {author} {\bibinfo {author} {\bibfnamefont {L.}~\bibnamefont
  {S\'ark\'any}}, \bibinfo {author} {\bibfnamefont {P.}~\bibnamefont {Weiss}},
  \bibinfo {author} {\bibfnamefont {H.}~\bibnamefont {Hattermann}}, \ and\
  \bibinfo {author} {\bibfnamefont {J.}~\bibnamefont {Fort\'agh}},\ }\href
  {\doibase 10.1103/PhysRevA.90.053416} {\bibfield  {journal} {\bibinfo
  {journal} {Phys. Rev. A}\ }\textbf {\bibinfo {volume} {90}},\ \bibinfo
  {pages} {053416} (\bibinfo {year} {2014})}\BibitemShut {NoStop}\bibitem [{\citenamefont {Hyafil}\ \emph {et~al.}(2004)\citenamefont {Hyafil},
  \citenamefont {Mozley}, \citenamefont {Perrin}, \citenamefont {Tailleur},
  \citenamefont {Nogues}, \citenamefont {Brune}, \citenamefont {Raimond},\ and\
  \citenamefont {Haroche}}]{hyafil:2004}  \BibitemOpen
  \bibfield  {author} {\bibinfo {author} {\bibfnamefont {P.}~\bibnamefont
  {Hyafil}}, \bibinfo {author} {\bibfnamefont {J.}~\bibnamefont {Mozley}},
  \bibinfo {author} {\bibfnamefont {A.}~\bibnamefont {Perrin}}, \bibinfo
  {author} {\bibfnamefont {J.}~\bibnamefont {Tailleur}}, \bibinfo {author}
  {\bibfnamefont {G.}~\bibnamefont {Nogues}}, \bibinfo {author} {\bibfnamefont
  {M.}~\bibnamefont {Brune}}, \bibinfo {author} {\bibfnamefont {J.~M.}\
  \bibnamefont {Raimond}}, \ and\ \bibinfo {author} {\bibfnamefont
  {S.}~\bibnamefont {Haroche}},\ }\href@noop {} {\bibfield  {journal} {\bibinfo
   {journal} {Phys. Rev. Lett.}\ }\textbf {\bibinfo {volume} {93}},\ \bibinfo
  {pages} {103001} (\bibinfo {year} {2004})}\BibitemShut {NoStop}\bibitem [{\citenamefont {Mozley}\ \emph {et~al.}(2005)\citenamefont {Mozley},
  \citenamefont {Hyafil}, \citenamefont {Nogues}, \citenamefont {Brune},
  \citenamefont {Raimond},\ and\ \citenamefont {Haroche}}]{mozley:2005}  \BibitemOpen
  \bibfield  {author} {\bibinfo {author} {\bibfnamefont {J.}~\bibnamefont
  {Mozley}}, \bibinfo {author} {\bibfnamefont {P.}~\bibnamefont {Hyafil}},
  \bibinfo {author} {\bibfnamefont {G.}~\bibnamefont {Nogues}}, \bibinfo
  {author} {\bibfnamefont {M.}~\bibnamefont {Brune}}, \bibinfo {author}
  {\bibfnamefont {J.-M.}\ \bibnamefont {Raimond}}, \ and\ \bibinfo {author}
  {\bibfnamefont {S.}~\bibnamefont {Haroche}},\ }\href@noop {} {\bibfield
  {journal} {\bibinfo  {journal} {Eur. Phys. J. D.}\ }\textbf {\bibinfo
  {volume} {35}},\ \bibinfo {pages} {43} (\bibinfo {year} {2005})}\BibitemShut
  {NoStop}\bibitem [{\citenamefont {Bason}\ \emph {et~al.}(2010)\citenamefont {Bason},
  \citenamefont {Tanasittikosol}, \citenamefont {Sargsyan}, \citenamefont
  {Mohapatra}, \citenamefont {Sarkisyan}, \citenamefont {Potvliege},\ and\
  \citenamefont {Adams}}]{bason:2010}  \BibitemOpen
  \bibfield  {author} {\bibinfo {author} {\bibfnamefont {M.~G.}\ \bibnamefont
  {Bason}}, \bibinfo {author} {\bibfnamefont {M.}~\bibnamefont
  {Tanasittikosol}}, \bibinfo {author} {\bibfnamefont {A.}~\bibnamefont
  {Sargsyan}}, \bibinfo {author} {\bibfnamefont {A.~K.}\ \bibnamefont
  {Mohapatra}}, \bibinfo {author} {\bibfnamefont {D.}~\bibnamefont
  {Sarkisyan}}, \bibinfo {author} {\bibfnamefont {R.~M.}\ \bibnamefont
  {Potvliege}}, \ and\ \bibinfo {author} {\bibfnamefont {C.~S.}\ \bibnamefont
  {Adams}},\ }\href@noop {} {\bibfield  {journal} {\bibinfo  {journal} {New J.
  Phys.}\ }\textbf {\bibinfo {volume} {12}},\ \bibinfo {pages} {065015}
  (\bibinfo {year} {2010})}\BibitemShut {NoStop}\bibitem [{\citenamefont {{Sevin\c cli}}\ and\ \citenamefont
  {Pohl}(2014)}]{sevincli:2014}  \BibitemOpen
  \bibfield  {author} {\bibinfo {author} {\bibfnamefont {S.}~\bibnamefont
  {{Sevin\c cli}}}\ and\ \bibinfo {author} {\bibfnamefont {T.}~\bibnamefont
  {Pohl}},\ }\href {http://stacks.iop.org/1367-2630/16/i=12/a=123036}
  {\bibfield  {journal} {\bibinfo  {journal} {New Journal of Physics}\ }\textbf
  {\bibinfo {volume} {16}},\ \bibinfo {pages} {123036} (\bibinfo {year}
  {2014})}\BibitemShut {NoStop}\bibitem [{\citenamefont {Jones}\ \emph {et~al.}(2013)\citenamefont {Jones},
  \citenamefont {Carter},\ and\ \citenamefont {Martin}}]{jones:2013}  \BibitemOpen
  \bibfield  {author} {\bibinfo {author} {\bibfnamefont {L.~A.}\ \bibnamefont
  {Jones}}, \bibinfo {author} {\bibfnamefont {J.~D.}\ \bibnamefont {Carter}}, \
  and\ \bibinfo {author} {\bibfnamefont {J.~D.~D.}\ \bibnamefont {Martin}},\
  }\href@noop {} {\bibfield  {journal} {\bibinfo  {journal} {Phys. Rev. A}\
  }\textbf {\bibinfo {volume} {87}},\ \bibinfo {pages} {023423} (\bibinfo
  {year} {2013})}\BibitemShut {NoStop}\bibitem [{\citenamefont {Hulet}\ and\ \citenamefont
  {Kleppner}(1983)}]{hulet:1983}  \BibitemOpen
  \bibfield  {author} {\bibinfo {author} {\bibfnamefont {R.~G.}\ \bibnamefont
  {Hulet}}\ and\ \bibinfo {author} {\bibfnamefont {D.}~\bibnamefont
  {Kleppner}},\ }\href {\doibase 10.1103/PhysRevLett.51.1430} {\bibfield
  {journal} {\bibinfo  {journal} {Phys. Rev. Lett.}\ }\textbf {\bibinfo
  {volume} {51}},\ \bibinfo {pages} {1430} (\bibinfo {year}
  {1983})}\BibitemShut {NoStop}\bibitem [{\citenamefont {Bethe}\ and\ \citenamefont
  {Salpeter}(1957)}]{bethe:1957}  \BibitemOpen
  \bibfield  {author} {\bibinfo {author} {\bibfnamefont {H.~A.}\ \bibnamefont
  {Bethe}}\ and\ \bibinfo {author} {\bibfnamefont {E.~E.}\ \bibnamefont
  {Salpeter}},\ }\href@noop {} {\emph {\bibinfo {title} {Quantum mechanics of
  one- and two-electron atoms}}}\ (\bibinfo  {publisher} {Springer-Verlag},\
  \bibinfo {address} {Berlin},\ \bibinfo {year} {1957})\BibitemShut {NoStop}\bibitem [{\citenamefont {Zimmerman}\ \emph {et~al.}(1979)\citenamefont
  {Zimmerman}, \citenamefont {Littman}, \citenamefont {Kash},\ and\
  \citenamefont {Kleppner}}]{zimmerman:1979}  \BibitemOpen
  \bibfield  {author} {\bibinfo {author} {\bibfnamefont {M.~L.}\ \bibnamefont
  {Zimmerman}}, \bibinfo {author} {\bibfnamefont {M.~G.}\ \bibnamefont
  {Littman}}, \bibinfo {author} {\bibfnamefont {M.~M.}\ \bibnamefont {Kash}}, \
  and\ \bibinfo {author} {\bibfnamefont {D.}~\bibnamefont {Kleppner}},\
  }\href@noop {} {\bibfield  {journal} {\bibinfo  {journal} {Phys. Rev. A}\
  }\textbf {\bibinfo {volume} {20}},\ \bibinfo {pages} {2251} (\bibinfo {year}
  {1979})}\BibitemShut {NoStop}\bibitem [{\citenamefont {Li}\ \emph {et~al.}(2003)\citenamefont {Li},
  \citenamefont {Mourachko}, \citenamefont {Noel},\ and\ \citenamefont
  {Gallagher}}]{li:2003}  \BibitemOpen
  \bibfield  {author} {\bibinfo {author} {\bibfnamefont {W.}~\bibnamefont
  {Li}}, \bibinfo {author} {\bibfnamefont {I.}~\bibnamefont {Mourachko}},
  \bibinfo {author} {\bibfnamefont {M.~W.}\ \bibnamefont {Noel}}, \ and\
  \bibinfo {author} {\bibfnamefont {T.~F.}\ \bibnamefont {Gallagher}},\
  }\href@noop {} {\bibfield  {journal} {\bibinfo  {journal} {Phys. Rev. A}\
  }\textbf {\bibinfo {volume} {67}},\ \bibinfo {pages} {052502} (\bibinfo
  {year} {2003})}\BibitemShut {NoStop}\bibitem [{\citenamefont {Eastham}(1973)}]{eastham:1973}  \BibitemOpen
  \bibfield  {author} {\bibinfo {author} {\bibfnamefont {M.}~\bibnamefont
  {Eastham}},\ }\href {https://books.google.ca/books?id=LUHvAAAAMAAJ} {\emph
  {\bibinfo {title} {The spectral theory of periodic differential
  equations}}},\ Texts in mathematics\ (\bibinfo  {publisher} {Scottish
  Academic Press [distributed by Chatto \& Windus, London]},\ \bibinfo {year}
  {1973})\BibitemShut {NoStop}\bibitem [{\citenamefont {Shirley}(1965)}]{shirley:1965}  \BibitemOpen
  \bibfield  {author} {\bibinfo {author} {\bibfnamefont {J.~H.}\ \bibnamefont
  {Shirley}},\ }\href@noop {} {\bibfield  {journal} {\bibinfo  {journal} {Phys.
  Rev.}\ }\textbf {\bibinfo {volume} {138}},\ \bibinfo {pages} {B979} (\bibinfo
  {year} {1965})}\BibitemShut {NoStop}\bibitem [{\citenamefont {van~de Water}\ \emph {et~al.}(1990)\citenamefont
  {van~de Water}, \citenamefont {Yoakum}, \citenamefont {van Leeuwen},
  \citenamefont {Sauer}, \citenamefont {Moorman}, \citenamefont {Galvez},
  \citenamefont {Mariani},\ and\ \citenamefont {Koch}}]{vandewater:1990}  \BibitemOpen
  \bibfield  {author} {\bibinfo {author} {\bibfnamefont {W.}~\bibnamefont
  {van~de Water}}, \bibinfo {author} {\bibfnamefont {S.}~\bibnamefont
  {Yoakum}}, \bibinfo {author} {\bibfnamefont {T.}~\bibnamefont {van Leeuwen}},
  \bibinfo {author} {\bibfnamefont {B.~E.}\ \bibnamefont {Sauer}}, \bibinfo
  {author} {\bibfnamefont {L.}~\bibnamefont {Moorman}}, \bibinfo {author}
  {\bibfnamefont {E.~J.}\ \bibnamefont {Galvez}}, \bibinfo {author}
  {\bibfnamefont {D.~R.}\ \bibnamefont {Mariani}}, \ and\ \bibinfo {author}
  {\bibfnamefont {P.~M.}\ \bibnamefont {Koch}},\ }\href {\doibase
  10.1103/PhysRevA.42.572} {\bibfield  {journal} {\bibinfo  {journal} {Phys.
  Rev. A}\ }\textbf {\bibinfo {volume} {42}},\ \bibinfo {pages} {572} (\bibinfo
  {year} {1990})}\BibitemShut {NoStop}\bibitem [{\citenamefont {Golub}\ and\ \citenamefont
  {Van~Loan}(2012)}]{golub:2012}  \BibitemOpen
  \bibfield  {author} {\bibinfo {author} {\bibfnamefont {G.}~\bibnamefont
  {Golub}}\ and\ \bibinfo {author} {\bibfnamefont {C.}~\bibnamefont
  {Van~Loan}},\ }\href {https://books.google.ca/books?id=X5YfsuCWpxMC} {\emph
  {\bibinfo {title} {Matrix Computations}}},\ Johns Hopkins Studies in the
  Mathematical Sciences\ (\bibinfo  {publisher} {Johns Hopkins University
  Press},\ \bibinfo {year} {2012})\BibitemShut {NoStop}\bibitem [{\citenamefont {Huby}(1961)}]{huby:1961}  \BibitemOpen
  \bibfield  {author} {\bibinfo {author} {\bibfnamefont {R.}~\bibnamefont
  {Huby}},\ }\href@noop {} {\bibfield  {journal} {\bibinfo  {journal} {Proc.
  Phys. Soc.}\ }\textbf {\bibinfo {volume} {78}},\ \bibinfo {pages} {529}
  (\bibinfo {year} {1961})}\BibitemShut {NoStop}\bibitem [{num()}]{numbers_for_dressed_nulling_theory:2015}  \BibitemOpen
  \href@noop {} {}\bibinfo {howpublished}
  {\url{https://github.com/jddmartin/numbers_for_dressed_nulling_theory}}\BibitemShut
  {NoStop}\bibitem [{\citenamefont {Ye}\ \emph {et~al.}(2008)\citenamefont {Ye},
  \citenamefont {Kimble},\ and\ \citenamefont {Katori}}]{ye:2008}  \BibitemOpen
  \bibfield  {author} {\bibinfo {author} {\bibfnamefont {J.}~\bibnamefont
  {Ye}}, \bibinfo {author} {\bibfnamefont {H.~J.}\ \bibnamefont {Kimble}}, \
  and\ \bibinfo {author} {\bibfnamefont {H.}~\bibnamefont {Katori}},\
  }\href@noop {} {\bibfield  {journal} {\bibinfo  {journal} {Science}\ }\textbf
  {\bibinfo {volume} {320}},\ \bibinfo {pages} {1734} (\bibinfo {year}
  {2008})}\BibitemShut {NoStop}\bibitem [{key(2011)}]{keysight:5990-7529EN}  \BibitemOpen
  \href@noop {} {\enquote {\bibinfo {title} {Reducing phase noise at {RF} and
  microwave frequencies, {K}eysight {T}echnologies {A}pplication {N}ote
  5990-7529},}\ } (\bibinfo {year} {2011})\BibitemShut {NoStop}\bibitem [{\citenamefont {Gross}\ and\ \citenamefont
  {Liang}(1986)}]{gross:1986}  \BibitemOpen
  \bibfield  {author} {\bibinfo {author} {\bibfnamefont {M.}~\bibnamefont
  {Gross}}\ and\ \bibinfo {author} {\bibfnamefont {J.}~\bibnamefont {Liang}},\
  }\href {\doibase 10.1103/PhysRevLett.57.3160} {\bibfield  {journal} {\bibinfo
   {journal} {Phys. Rev. Lett.}\ }\textbf {\bibinfo {volume} {57}},\ \bibinfo
  {pages} {3160} (\bibinfo {year} {1986})}\BibitemShut {NoStop}\bibitem [{\citenamefont {Brune}\ \emph {et~al.}(1996)\citenamefont {Brune},
  \citenamefont {Schmidt-Kaler}, \citenamefont {Maali}, \citenamefont {Dreyer},
  \citenamefont {Hagley}, \citenamefont {Raimond},\ and\ \citenamefont
  {Haroche}}]{brune:1999}  \BibitemOpen
  \bibfield  {author} {\bibinfo {author} {\bibfnamefont {M.}~\bibnamefont
  {Brune}}, \bibinfo {author} {\bibfnamefont {F.}~\bibnamefont
  {Schmidt-Kaler}}, \bibinfo {author} {\bibfnamefont {A.}~\bibnamefont
  {Maali}}, \bibinfo {author} {\bibfnamefont {J.}~\bibnamefont {Dreyer}},
  \bibinfo {author} {\bibfnamefont {E.}~\bibnamefont {Hagley}}, \bibinfo
  {author} {\bibfnamefont {J.~M.}\ \bibnamefont {Raimond}}, \ and\ \bibinfo
  {author} {\bibfnamefont {S.}~\bibnamefont {Haroche}},\ }\href {\doibase
  10.1103/PhysRevLett.76.1800} {\bibfield  {journal} {\bibinfo  {journal}
  {Phys. Rev. Lett.}\ }\textbf {\bibinfo {volume} {76}},\ \bibinfo {pages}
  {1800} (\bibinfo {year} {1996})}\BibitemShut {NoStop}\bibitem [{\citenamefont {Carter}\ and\ \citenamefont
  {Martin}(2011)}]{carter:2011}  \BibitemOpen
  \bibfield  {author} {\bibinfo {author} {\bibfnamefont {J.~D.}\ \bibnamefont
  {Carter}}\ and\ \bibinfo {author} {\bibfnamefont {J.~D.~D.}\ \bibnamefont
  {Martin}},\ }\href@noop {} {\bibfield  {journal} {\bibinfo  {journal} {Phys.
  Rev. A}\ }\textbf {\bibinfo {volume} {83}},\ \bibinfo {pages} {032902}
  (\bibinfo {year} {2011})}\BibitemShut {NoStop}\bibitem [{\citenamefont {Schindler}\ \emph {et~al.}(2015)\citenamefont
  {Schindler}, \citenamefont {Gorman}, \citenamefont {Daniilidis},\ and\
  \citenamefont {H\"affner}}]{schindler:2015}  \BibitemOpen
  \bibfield  {author} {\bibinfo {author} {\bibfnamefont {P.}~\bibnamefont
  {Schindler}}, \bibinfo {author} {\bibfnamefont {D.~J.}\ \bibnamefont
  {Gorman}}, \bibinfo {author} {\bibfnamefont {N.}~\bibnamefont {Daniilidis}},
  \ and\ \bibinfo {author} {\bibfnamefont {H.}~\bibnamefont {H\"affner}},\
  }\href {\doibase 10.1103/PhysRevA.92.013414} {\bibfield  {journal} {\bibinfo
  {journal} {Phys. Rev. A}\ }\textbf {\bibinfo {volume} {92}},\ \bibinfo
  {pages} {013414} (\bibinfo {year} {2015})}\BibitemShut {NoStop}\bibitem [{\citenamefont {Mozley}(2004)}]{mozley:2004}  \BibitemOpen
  \bibfield  {author} {\bibinfo {author} {\bibfnamefont {J.}~\bibnamefont
  {Mozley}},\ }\emph {\bibinfo {title} {Micro-pi\'{e}geage des atomes de
  {R}ubidium dans un environnement cryog\'{e}nique}},\ \href@noop {} {Ph.D.
  thesis},\ \bibinfo  {school} {Universit\'{e} Paris VI}, \bibinfo {address}
  {Paris} (\bibinfo {year} {2004})\BibitemShut {NoStop}\bibitem [{\citenamefont {Park}(1960)}]{park:1960}  \BibitemOpen
  \bibfield  {author} {\bibinfo {author} {\bibfnamefont {D.}~\bibnamefont
  {Park}},\ }\href@noop {} {\bibfield  {journal} {\bibinfo  {journal}
  {Zeitschrift f\"{u}r Physik}\ }\textbf {\bibinfo {volume} {159}},\ \bibinfo
  {pages} {155} (\bibinfo {year} {1960})}\BibitemShut {NoStop}\end{thebibliography}
\end{document}